\documentclass[letterpaper,twocolumn,english,preprintnumbers,amsmath,amssymb,superscriptaddress,footinbib,prx]{revtex4-2}
\usepackage{graphicx}

\graphicspath{{./Images/}}
\usepackage{xcolor}
\usepackage[colorlinks = true, citecolor=blue, linkcolor=black,urlcolor=purple]{hyperref}
\usepackage{enumitem}
    \setlist[itemize]{noitemsep, topsep=0pt}
    \setlist[enumerate]{noitemsep, topsep=0pt}
\usepackage{verbatim}
\usepackage{mathtools,amssymb,amsmath}
\usepackage{bm}
\usepackage{yhmath}
\usepackage[inline]{asymptote}
\usepackage{cancel}
\usepackage{relsize}
\usepackage{array}
\usepackage{bm}
\newtheorem{thm}{Theorem}

\DeclareMathOperator{\Var}{Var}

\newcommand{\be}{\begin{equation}}
\newcommand{\ee}{\end{equation}}
\newcommand{\<}{\langle}
\renewcommand{\>}{\rangle}

\newcommand{\Tr}{{\rm Tr\,}}

\usepackage[normalem]{ulem}
\usepackage{color}

\usepackage{algorithm}
\usepackage{algpseudocode}

\renewcommand{\vec}[1]{{\bf #1}}
\newtheorem{theoremS}{Theorem}

\newtheorem{lemmaS}{Lemma}

\makeatletter
\newsavebox{\@brx}
\newcommand{\llangle}[1][]{\savebox{\@brx}{\(\m@th{#1\langle}\)}
  \mathopen{\copy\@brx\kern-0.5\wd\@brx\usebox{\@brx}}}
\newcommand{\rrangle}[1][]{\savebox{\@brx}{\(\m@th{#1\rangle}\)}
  \mathclose{\copy\@brx\kern-0.5\wd\@brx\usebox{\@brx}}}
\makeatother

\newcommand{\almaden}{IBM Quantum, IBM Research -- Almaden, San Jose CA, 95120, USA}
\newcommand{\cambridge}{IBM Quantum, MIT-IBM Watson AI lab,  Cambridge MA, 02142, USA}
\newcommand{\google}{Google Quantum AI, Venice Beach, CA 90291, USA}

\newcommand{\figsize}{1}
\newcommand{\figsizenew}{1}

\begin{document}
\makeatother

\title{Preparing thermal states on noiseless and noisy\\ programmable quantum processors} 

\author{Oles Shtanko}
\email{oles.shtanko@ibm.com}

\affiliation{\almaden}

\author{Ramis Movassagh}
\email{movassagh@google.com}

\affiliation{\cambridge}
\affiliation{\google}

\begin{abstract}
Nature is governed by precise physical laws, which can inspire the discovery of new computer-run simulation algorithms.
Thermal states are the most ubiquitous for they are the equilibrium states of matter. 
Simulating thermal states of quantum matter has applications ranging from quantum machine learning to better understanding of high-temperature superconductivity and quantum chemistry. The computational complexity of this task is hopelessly hard for classical computers \cite{sign1990loh}. The existing quantum algorithms come with caveats: most either require quantum phase estimation \cite{poulin2009,riera2012,Temme2011,Yung2012} rendering them impractical for current noisy hardware, or are variational~\cite{Cerezo2021,wu2019,verdon2019quantum} which face obstacles such as initialization, barren plateaus, and a general lack of provable guarantee~\cite{McClean2018}. We provide two quantum algorithms with provable guarantees to prepare thermal states on (near-term) quantum computers that avoid these drawbacks. 
The first algorithm is inspired by the natural thermalization process where the ancilla qubits act as the infinite thermal bath.
This algorithm can potentially run in polynomial time to sample thermal distributions of ergodic systems-- the vast class of physical systems that equilibrate in isolation with respect to local observables. 
The second algorithm works for any system and in general runs in exponential time. However, it requires significantly smaller quantum resources than previous such algorithms. 
In addition, we provide an error mitigation technique for both algorithms to fight back decoherence, which enables us to run our algorithms on the near-term quantum devices. To illustration, we simulate the thermal state of the hardcore Bose-Hubbard model on the latest generation of available quantum computers.
\end{abstract}

\maketitle

Nature has inspired many modern classical algorithms \cite{brabazon2015natural}. Here, we propose a new algorithm inspired by natural equilibration that is designed to prepare Gibbs states on quantum hardware.

Gibbs states are fundamental as they represent the natural equilibrium state of many-body systems in physics and chemistry. Simulating Gibbs distributions on a quantum hardware offers an opportunity to explore equilibrium many-body phenomena \cite{georgescu2014quantum}, including strongly-correlated quantum matter \cite{hofstetter2018quantum} and high-energy physics \cite{wiese2013ultracold}. The quantum simulation of Gibbs states would enable accurate predictions of molecular structures and reaction rates \cite{Bauer2020}. As such, it provides a significant advantage over classical ones as they suffer from the infamous sign problem \cite{sign1990loh}. The Gibbs states are used in computer science and quantum computing as well. For example, the low-temperature Gibbs states provide an approximation to optimization problems by encoding them into the lowest energy states \cite{somma2008quantum}. In quantum machine learning, Gibbs distribution can be used to create generative models that are superior to classical counterparts \cite{amin2018quantum}. Finally, for systems with sufficiently large energy gap between the smallest two energies (so-called finite energy gap), preparation of Gibbs states may recover the quantum information  encoded in the ground state \cite{brown2016quantum}. Thus, there is an active frontier for developing heuristic methods to simulate thermalization tailored to different experimental platforms \cite{Shabani2016,Ashida2018,Su2020,metcalf2020,Schulman2005,Kapit2015,allahverdyan2004}.
 
Preparation of generic Gibbs states is a challenging task from the standpoint of computational complexity. For classical systems, sampling from the Gibbs state is a NP-hard problem. In the quantum case, this problem is believed to be even harder and be complete in the formidable complexity class known as quantum Merlin Arthur (QMA) \cite{hallgren2013local,Aharonov2009,Schuch2007,Piddock2015,Bausch2017}, which is the analog of NP for quantum computers. It is therefore unlikely that an efficient algorithm can be found that is applicable to all quantum systems.  In fact algorithms with provable performance  \cite{poulin2009,riera2012,biglin2010,chowdhury2016quantum} have  convergence times that scale exponentially with the system size. However, some empirical algorithms, such as the quantum Metropolis algorithm \cite{Temme2011,Yung2012,Moussa2019}, can succeed in polynomial time for certain systems. 

A formal implementation of quantum MH algorithm utilizes non-local operations as well as the quantum phase estimation subroutine, which is resource-intensive and very sensitive to noise rendering it impractical especially in the near-term. Despite recent progress in hardware-efficient algorithms that use only local gates and designed for low-correlated systems \cite{Motta2020,Sun2021}, there is push for constructing Gibbs states for more general classes of systems. 

The quantum dynamics of an isolated system is generated by its Hamiltonian.  Given the target Hamiltonian $H$, the goal is to obtain samples from the mixed state
\be
\rho_\beta = \frac{1}{Z_\beta}\exp(-\beta H), \quad Z_\beta = \Tr \left[\exp(-\beta H)\right],
\ee
where $\beta$ is inverse temperature and $Z_\beta$ is a partition function. This state is the fixed point of the evolution of the system in contact with a much larger (hypothetically infinite) reservoir at inverse temperature $\beta$.

In this paper, we present two algorithms for preparing the Gibbs state on the quantum computer that avoid the quantum phase estimation subroutine. The first algorithm is inspired by the natural thermalization process where the ancilla qubits act as the infinite thermal bath~\cite{terhal2000,Su2020,polla2021}.
This algorithm can potentially run in polynomial time to sample thermal distributions of ergodic systems-- the vast class of physical systems that equilibrate in isolation with respect to local observables. 
The second algorithm works for any system and in general runs in exponential time. However, it requires significantly smaller quantum resources than previous such algorithms. Both algorithms can be run on the near-term noisy quantum computers as we will demonstrate below.

Our algorithms have the advantage of utilizing random local quantum circuits, which  by definition are not fine-tuned. Motivated by the challenges of the near-term quantum computation era in which effects of noise and decoherence cannot yet be corrected, we find that treating the random parameters in the circuits as adjustable leads to substantial mitigation of errors caused by decoherence and noise. The adjustment of parameters resembles that of the  variational algorithms \cite{Cerezo2021,wu2019,verdon2019quantum}. However, in this work the optimization over the parameters only serves to reduce the effects of noise. Consequently, the algorithms do not suffer as much from the problem of barren plateaus \cite{McClean2018}.\\

\textbf{Quantum Ergodicity}. 
In Nature one encounters two major classes of physical systems that in some ways are at extremes. These are integrable and chaotic systems.  

Integrable systems are often idealizations constrained by symmetries, transition rules, and integrals of motion. They may allow for exact analytical solutions such as those obtained by the Yang-Baxter equations or Bethe-Ansatz.  Chaotic systems on the other hand lack the fine tuning necessary for integrability. These include most  interacting systems in presence of weak disorder.  Chaotic systems derive significance from the fact that the behavior of weakly perturbed models  are well-approximated by chaotic behavior at sufficiently high temperatures. 

The dynamics of chaotic systems are ergodic, which means they are only constrained by energy conservation.
A key feature of ergodic systems is that their local observables usually quickly reach their thermal equilibrium values. Calculating these thermal values is often challenging for traditional analytical or numerical methods. However, as we will show, it is possible to use quantum computers to generate quantum states that accurately describe the thermal state of an ergodic system. The behavior of an ergodic system is commonly described by ETH
which are certain conditions on the eigenstates $|\mu\>$ of the Hamiltonian 
 of the system~\cite{deutch1991quantum,srednicki1999approach,srednicki1994chaos,d2016quantum,deutsch2018eigenstate,d2016quantum,dymarsky2022}.

Mathematically, ETH's premise is that
for any \textit{local} operator $V$  and any two eigenstates $|\mu\>$ and $|\nu\>$ within a certain energy interval, the matrix elements $\<\mu|V|\nu\>$ satisfy 
$\<\mu|V|\nu\> = v_\mu\delta_{\mu\nu}+
e^{-O(n)}R_{\mu\nu}$, 
where
$v_\mu$ are real numbers, $n\gg1$ is the number of qubits, and  $R_{\mu\nu}$ are standard Gaussian variables.
In words, ETH says that the off-diagonal elements are random and  exponentially small in $n$. This property has been numerically verified for a wide range of physical systems ~\cite{rigol2008thermalization,d2016quantum,deutsch2018eigenstate}.

Our ergodic algorithm described just below gives rigorous guarantees for the preperation of the thermal state on a quantum computer for systems that obey the ETH. We leave the question of the performance of this algorithm for more general Hamiltonians for future work. We will show that the infinite reservoir is mimicked by a finite number of independent, properly reset and randomized ancilla-qubits, that are weakly coupled to the system. The ancilla qubits extract entropy from the system qubits in the course of the quantum  computation, whereby thermalize the system-qubits at inverse temperature $\beta$. This essentially a quantum circuit adaption of the textbook concept of bringing the system in contact with an infinite reservoir, where the reservoir is played by the ancilla qubits.\\

\begin{figure}[t]
 \centering
  \includegraphics[width=\figsize\linewidth]{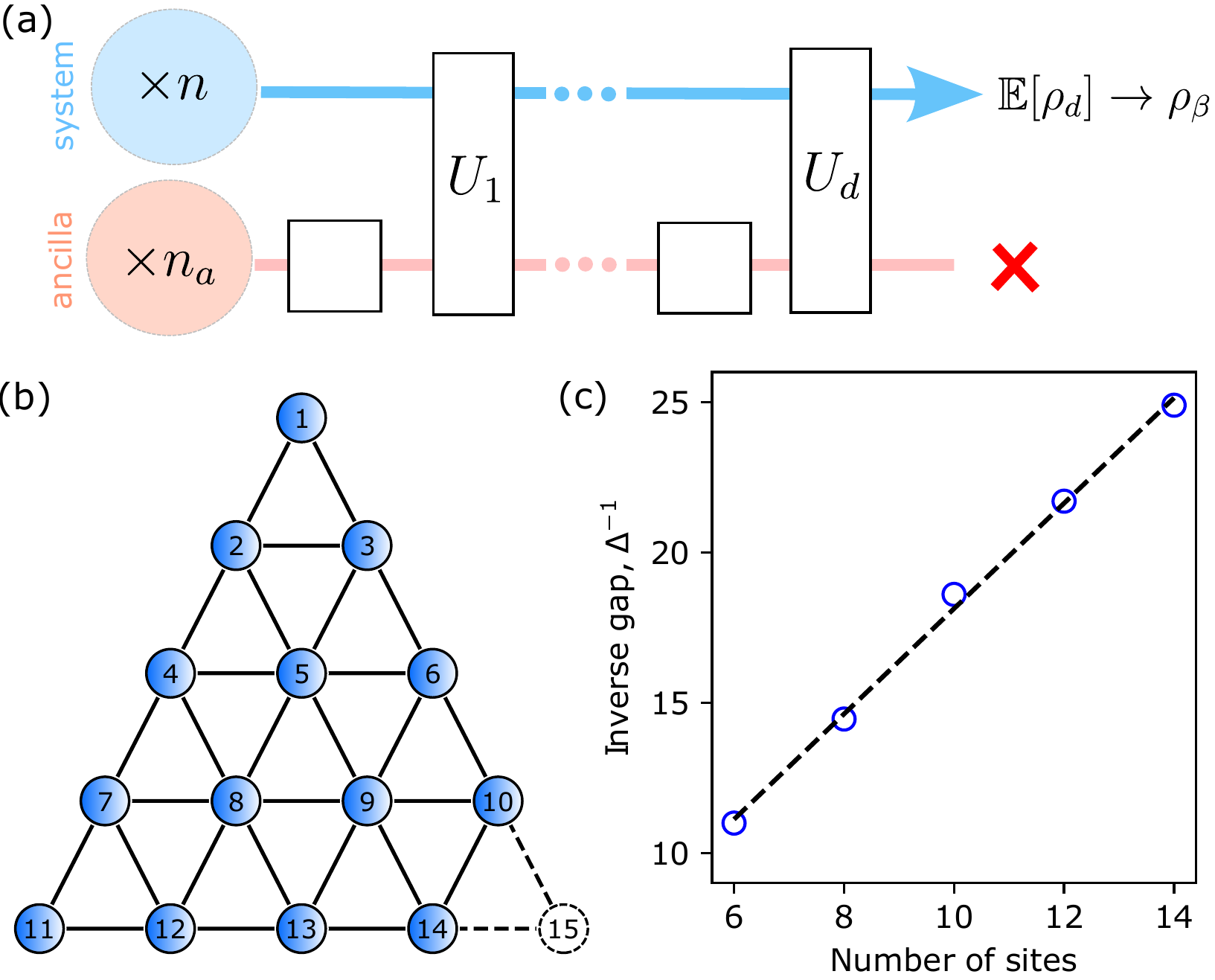}
  \caption{\textbf{Ergodic algorithm and its performance} (a) Illustration of the ergodic algorithm. The qubits are divided into two parts, a system of $n$ qubits and an ancilla of $n_a$ qubits. The ancilla is reset to a thermal state at the beginning of each cycle; during the cycle both parts are weakly coupled by the Hamiltonian in Eq.~\eqref{eq:methdo2_ham}. (b) The layout of a lattice half-filled with hardcore bosons obeying the Hamiltonian in Eq.~\eqref{eq:hamiltonian} with $J=1$ and $U=0.1$. The circles represent the sites that can be occupied by a single boson, the edges connect the nearest neighbors. (c) The inverse gap $\Delta^{-1}$  of the underlying classical Markov process in our algorithm (Theorem \ref{eq:ergodic_theorem2}) for the system in (b) with $\beta = 1$, $\Omega=1$, and $\gamma=0.1$. The inverse gap grows linearly with the system size of the truncated system (as shown by the dashed line), where the truncation is performed in numerical order (i.e., a system of $n$ sites includes the sites $1,\dots,n$).}
  \label{fig:fig1}
\end{figure}

\textbf{Ergodic algorithm}.
The summary of the algorithm is as follows. First, we assign $n$ qubits to be the ``system" qubits on which the target Hamiltonian $H$ acts. We initialize these to a random product state. We then pick $n_a$ ``ancilla" qubits and initialize each to $|0\rangle$. The algorithm consists of $d$ cycles (Fig.~\ref{fig:fig1}a). At the beginning of each cycle, the ``ancilla" qubits are reset to $|0\rangle$ and then each is easily put in a single-qubit thermal equilibrium state of an assigned random single-qubit Hamiltonian. Throughout the cycle, all the qubits evolve under combined Hamiltonians of the ancilla and the system with a weak coupling between them (see Fig.~\ref{fig:fig1}b). The evolution time at the $k$th cycle, $ t_k$, is random from the exponential  distribution $\gamma\exp(-\gamma t_k)$, and is generated by the Hamiltonian
\be\label{eq:methdo2_ham}
\mathcal H_k = H\otimes I_{\rm anc}+I_{\rm sys}\otimes H^k_{\rm anc}+\lambda \sum_{m=1}^{n_a} V_{km}\otimes X_m,
\ee
where $V_{km}$ are $\ell$-local random operators acting on the system qubits, $H^k_{\rm anc} = \frac 12 \sum_{m=1}^{n_a}\omega_{km} Z_m$ is a one-local Hamiltonian on the ancilla qubits, $X_m$ and $Z_m$ are a Pauli-X and Z on the $m$th ancilla qubit, $\omega_{km}\in [-\Omega,\Omega]$ is uniformly random, and $\lambda$ is the strength of the coupling. The process runs in time of $t=d/\gamma$ on average, where $d$ is the total number of cycles.

 We consider the regime where the parameters $\lambda$ and $\gamma$ are sufficiently small to allow for a weak coupling between the system and the ancilla. In this case, the expected density matrix is well described by a classical update rule given by a transition matrix $T$ for a choice of parameters.
 Thus, we bridge the quantum problem to the classical Markov process. The convergence of the process depends on the gap $\Delta$, i.e. the difference between the two largest eigenvalues of the corresponding matrix $T$. Under the above choice of parameters $\lambda,\gamma$, we obtain our main theorem (see Theorem~\ref{ergodic_theorem_detailed} in Supplementary Information for a formal statement):
 \begin{thm} \label{eq:ergodic_theorem2} Given an inverse temperature $\beta$, and the error tolerance $0<\epsilon\le1$, the expected output of the circuit $\mathbb{E}[\rho_d]$ becomes $\epsilon-$close to the true Gibbs state,
\[
 \| \mathbb{E}[\rho_d]-\rho_\beta\|_1\leq\epsilon,
\]
in time $t = \tilde O(\beta m^3/\epsilon^2)$, where $m \leq O(n/\Delta)$ is a mixing time of the corresponding Markov process.
\label{thm:2}
\end{thm}

Here $\tilde{O}$ hides poly-log factors. We note that $O(n/\Delta)$ is only an upper-bound on the mixing time, and the algorithm may run faster than $t = \tilde O(\beta n^3/\Delta^{3}\epsilon^2)$.

From this result we can conclude that if $\Delta$ is polynomially small in the system size, the algorithm converges in polynomial time as well. As an example, we analyze the gap for a finite ergodic system of hardcore bosons (see details of the Hamiltonian below). For sizes up to $14$ sites, this system exhibits a reasonable linear scaling of the inverse gap $\Delta^{-1}$ with the size of the system (Fig.~\ref{fig:fig1}c). 
A comparison of this algorithm to the well-known quantum Metropolis algorithm~\cite{Temme2011} can be found in the Supplementary Section \ref{Supp2:comparison Metropolis}. The resource estimation for this algorithm is shown in section \ref{Supp2:ResourceEstim_Erog}.

\begin{figure*}[t]
    \centering
  \includegraphics[width=1\linewidth]{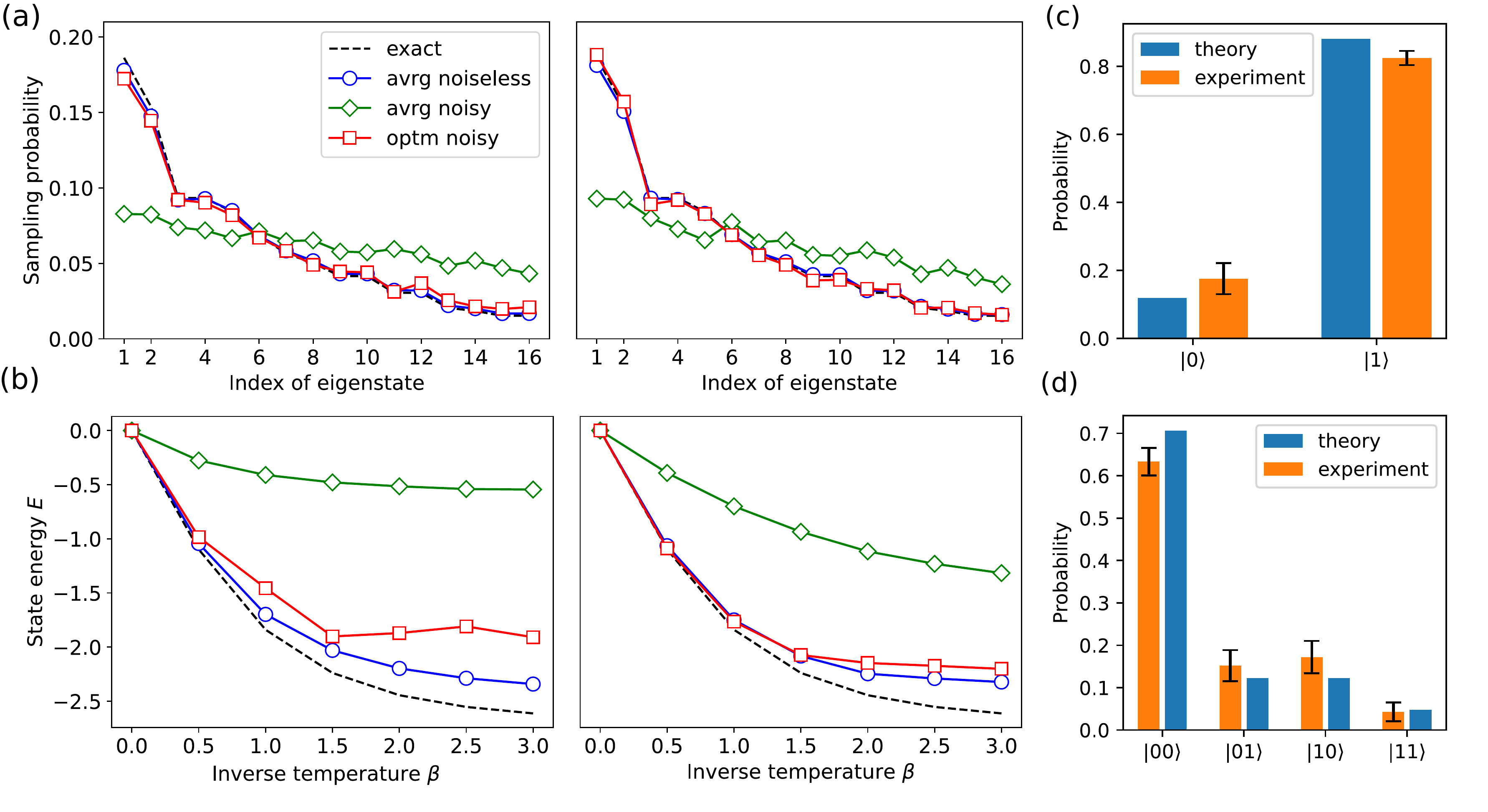}
\caption{\textbf{Simulations and experiments.} (a) and (b). Numerical simulations of a seven-qubit device including $n=4$  one-dimensional system qubits and $n_a = 3$ ancilla qubits for the target Hamiltonian in Eq.~\eqref{eq:hamiltonian}, where $J = U = 1$ .  Panel (a) shows the probabilities of sampling Hamiltonian eigenstates for ergodic algorithm (left) and universal algorithm (right). The curves describe the $\beta = 1$ output averaged over $10^3$ samples for: noiseless circuits (blue circles), noisy circuits (green diamonds), and  noise-optimized circuits (red squares). Dashed lines show the exact solution. The number of cycles in the Ergodic algorithm is $d=20$, $\gamma = 0.1$, and noise is modeled by single-qubit depolarizing channels after each cycle with the probability $p = \Gamma t$, where $t$ is the cycle time and $\Gamma = 10^{-3}$ is the noise rate. In the universal algorithm, we take $d=5$ cycles and take depolarizing noise which affects each qubit with probability $p =10^{-2}$ after 2-qubit gates and $p'=2\cdot 10^{-2}$ after 3-qubit gates (see Supplementary Section~\ref{Supp1:F} for justification). (b) The expected energy of the output as the function of temperature in the same setting.  (c) and (d) Implementation on IBM 7-qubit \textit{ibm\_casablanca} device. Histograms show the experimental sampling probabilities compared with the theoretical predictions. (c) Implementation of the ergodic algorithm for $n=1$ system qubit and $n_a = 1$ ancilla qubit for Hamiltonian $H = Z_1$. The coupling is $\lambda = 0.1$ for the first cycle and it decreases linearly with number of cycles, inverse average time is $\gamma = 0.01$. Experiment uses $150$ random circuit configurations with $8192$ samples per circuit. (d)  Sampling probabilities for the universal algorithm utilizing $n_a = 3$ ancilla qubits and $n=2$ system qubits with the Hamiltonian $H = X_1X_2-Z_1-Z_2$.  In the experiment  we took $100$ random circuit configurations with $8192$ samples per circuit.}
  \label{fig:fig2}
\end{figure*}
Even though ergodic systems are common in nature, one may require a method for preparing Gibbs states for \textit{general} Hamiltonians.
Unlike in the ergodic setting, by to the QMA-hardness of Gibbs sampling for general Hamiltonians~\cite{hallgren2013local,Aharonov2009,Schuch2007,Piddock2015,Bausch2017},  it is expected that the scaling of the run-time for a universal algorithm would be exponential.  Our second algorithm utilizes monitored random quantum circuits to prepare the Gibbs state. Random quantum circuits are also the backbone of models of quantum chaos \cite{zhou2020entanglement}, dynamic quantum phases \cite{lavasani2021measurement}, and modern quantum complexity theory \cite{movassagh2019quantum,kondo2021fine,bouland2019complexity}. We will show that any realization of our random circuit is likely a good Gibbs sampler, whereas averaging over circuits provides significant improvement compared to individual realizations.\\

\textbf{Universal algorithm}.
In simple terms, this algorithm can be understood as applying an effective imaginary-time evolution operator $V_\beta \propto e^{-\beta H/2}$. Consider starting from a random initial state in the computational basis $|\vec z\> = |z_1\dots z_n\>$, where $z_i\in\{0,1\}$. Then the average over the initial state  density matrix after the imaginary-time evolution is
\be
\mathbb E_{\vec z} V_\beta|\vec z\>\<\vec z|V^\dag_\beta \propto \exp(-\beta H)\propto\rho_\beta.
\ee
To implement this operation, we construct a circuit that utilizes ancilla and intermediate measurements \cite{chowdhury2016quantum,liu2021probabilistic}.

For simplicity, consider Hamiltonians of the form $H = \sum_{m=1}^M h_m$, where $h_m\ge0$ are positive local terms.  Given a depth $d$, we define a small imaginary-time step $\delta\beta = \beta/d$, and the set of angles $\theta_{mk}\in \mathcal N(0,1)$ to be standard normals. For each cycle $k$ and term $h_m$ we initialize an ancilla qubit to $|0\rangle$, and apply the gate
\be\label{eq:rqc_gate}
U(\theta_{mk},h_m) = \exp(i \theta_{mk} \sqrt{\delta\beta h_m}\otimes X)
\ee
to the system qubits in the support of $h_m$ and the ancilla qubit, where $X$ is a Pauli-$X$ acting on the  ancilla. We then measure the corresponding ancilla qubit. At the end, we accept the output of the circuit only if all of the ancilla qubits evaluate to $0$, otherwise we re-run the circuit. This  is the source of exponential slow-down mentioned above. A gaussian integration reveals that the average over $\theta_{mk}$ results in each gate essentially enacting a short imaginary time evolution for $h_m$. Therefore, the overall imaginary time of  evolution, $\beta$, is composed from such gates in a trotterized manner \cite{childs2021theory} in $d$ cycles. Fig.~\ref{fig:fig1}(b) shows an example of the $k$th cycle.

The following theorem quantifies the convergence of the output $\rho_{\rm out}$ to the Gibbs state:

 \begin{thm} \label{thm:main_text_universal} Let $\xi : = \beta^2 M/d\ll 1$.
 Then for any $0<\epsilon\leq1$, the output  after $d$ cycles satisfies
\begin{align}\label{eq:thm2_first_line}
&S(\rho_\beta\,\|\,\mathbb E[\rho_{\rm out}]) \leq  A\xi^2\;,\frac{}{}\\ 
&{\rm Prob}\Bigl(S(\rho_\beta\,\|\,\rho_{\rm out}) \geq \frac{\xi C}{\epsilon }\Bigl)\leq \epsilon \label{thm2_second_line},
\end{align}
where $S(\rho\|\sigma)$ is the relative entropy, and $A$ and $C$ are constants depending on $h_m$.\label{thm:1}
\end{thm}
The probability and expectation in this theorem are taken with respect to the distribution of random angles $\theta_{mk}$ and $\rho_{\rm out}$ is output of a successful circuit run (i.e., where all ancilla measurements return 0). 

To illustrate, let us assume that $\|h_m\|\leq O(1)$ and $M\propto n$, where $n$ is the number of qubits as before. Then, in order to achieve $S(\sigma\|\rho)\le \epsilon$ for a fixed error $\epsilon>0$, the minimal required circuit depth is $d\propto O(n\beta^2/\sqrt{\epsilon})$ for the averaged result in Eq.~\eqref{eq:thm2_first_line}, while a typical circuit requires depth $d\propto O(n\beta^2/\epsilon)$ according to Eq.~\eqref{thm2_second_line}.
 The acceptance probability for a large-$d$ circuit is $p \simeq 2^{-n} \mathcal Z_\beta = 2^{-\alpha n}$, where $\alpha =  1-(\log_2 \mathcal Z_\beta)/n$, therefore the algorithm time $t\propto 1/p$ grows exponentially with the number of qubits. 
 
 Due to the QMA-hardness of Gibbs sampling Hamiltonians, it is unlikely that this scaling can be improved without restrictions on the Hamiltonian.
 We detail the resource estimation for this algorithm in the Supplementary Section \ref{Supp2:ResourceEstim_Univ}.  Despite exponential scaling of the performance time, this algorithm provides sufficient improvement over comparable algorithms targeted general Hamiltonians \cite{poulin2009,riera2012,biglin2010,chowdhury2016quantum}.

We would like to remark on the utility of this algorithm when the terms $h_m$ are proportional to local Pauli operators. Then, for the single-qubit terms, the gate in Eq.~\eqref{eq:rqc_gate} can be implemented using only a single CNOT gate, compared to three CNOTs required for generic two-qubit gate \cite{zhang2003exact}. Moreover, a gate corresponding to a two system-qubits term involves the total of three qubits. This in our algorithm requires only two CNOTs, and is much less than what is required for a general three-qubit gate (see Supplementary Section~\ref{Supp1:F} for details).\\

\textbf{Demonstration}. 
We illustrate our algorithm for the hardcore Bose-Hubbard model, which is often used to study strong interactions in quantum many-body systems. The Hamiltonian of this model is
\be\label{eq:hamiltonian}
H = -J\sum_{\<i,j\>}(a^\dag_ia_{j}+a^\dag_{j}a_i)+U\sum_{\<i,j\>} n_i n_{j},
\ee
where $a_i$ are hard-core boson Fock operators satisfying $[a_j,a^\dag_j] = 0$ for all $i\neq j$ and $\{a_i,a^\dag_i\}=1$ at the same site, $n_i = a_i^\dag a_i$ are density operators, $J$ is the hopping coefficient, $U$ is the density-density coupling, and $\<i,j\>$ denotes the sum over nearest sites. This model can be mapped onto a qubit spin system (see Supplementary Section \ref{s_sec:fh_mapping}). The numerical behavior of the gap shown in Fig.~\ref{fig:fig1}c and the Theorem~\ref{eq:ergodic_theorem2} together would imply that one can prepare the Gibbs state of this model on a quantum computer to the statistical distance $\epsilon$ in time $t =O(\beta n^6/\epsilon^2)$.

As a test of the algorithms in realistic setting, we compare the probability of sampling Hamiltonian eigenstates for a noisy and noiseless limited-depth algorithms with corresponding Gibbs distribution, see Fig.~\ref{fig:fig2} (a) and (b).  In particular, the system is not exactly ergodic for these sizes, yet the algorithm successfully approximates Gibbs states. Also, numerical simulations provide evidence that the noise degrades the accuracy of the output distribution, but does not completely destroy the result in small systems. Another method of measuring the accuracy of the algorithm is comparison of the temperature dependency of the output state energy $E = \mathbb E \Tr(\rho_{\rm out}H)$ with the exact values $E_\beta =\Tr(\rho_\beta H)$. This comparison is shown in bottom panel of Fig.~\ref{fig:fig2} (a) and (b). As we analytically predicted above, the algorithm better performs at higher temperatures $\beta\ll J,U$. We also implement the algorithms on the latest version of IBM quantum hardware, as shown in Fig~\ref{fig:fig2} (c) and (d).\\

\textbf{Error mitigation}. Decoherence caused by noise is a major limitation for the current generation of quantum hardware. In addition to existing error mitigation techniques, here we show that both of our algorithms allow for a complementary way of error mitigation. We note that circuits in both algorithms depend on multiple random parameters. The choice of these parameters is mostly irrelevant in the ideal setting of zero noise and unlimited depth. Numerical results below show that, in a realistic setting, certain choices of parameters lead to a better performance, by
reducing the effect of noise. The result can be optimized over these parameters in order to significantly improve the results.

In order to improve the output of a noisy circuit, we take as the objective function the free energy of the output, and try to minimize it by adjusting the parameters in the model. We do not take into account the complexity of measuring free energy and leave it for future work. We find that such an optimization significantly reduces the effect of depolarizing noise, see Fig.~\ref{fig:fig2} (a) and (b).\\

\textbf{Discussions}. 
We provide two algorithms for sampling from the Gibbs distributions of quantum Hamiltonians. The first assumes ergodic Hamiltonians that obey the ETH hypothesis, and the second is universal. In the first case, we obtain a quantum algorithm that is implemented on a quantum circuit with local architecture whose complexity is similar to the quantum Metropolis-Hastings algorithms but does {\it not} require the quantum phase estimation subroutine. Our second algorithm can also be implemented on a quantum circuit and is in general an exponential time algorithm as expected.  We demonstrated both algorithms on actual near-term quantum hardware. In addition, we found a way to mitigate errors in both algorithms by running an optimization routine over the parameter space that define the gates of the circuits. We find that our error mitigation technique is quite effective and in some cases reduces the effect of noise almost completely.  Although, in general, error mitigation seems to require exponential samples \cite{takagi2022universal}, it remains an open problem whether for classes of Hamiltonians such as those obeying ETH more sample-efficient mitigation techniques exist.

Together with ours a recent simultaneous progress appeared \cite{BrandaoComment2021}. Our result for the ergodic algorithm can be combined with this recent work to show that the  QMH algorithms converge to the Gibbs state in a polynomial time for ETH systems \cite{chen2021fast}. This implies that ours is also a polynomial time algorithm. A further interesting aspect of the ergodic algorithm is the investigation of its applicability to non-equilibrium states of matter realized by Floquet Hamiltonians~\cite{khemani2016phase,bunin2011universal}.

A major open problem that is substantiated with our empirical explorations is to prove our first algorithm without the ETH assumption and for a more general class of Hamiltonians. 

Another interesting open problem is to use our techniques to prove the cooling rate into the ground state. One can envisage ancilla being prepared in their ground state and derive convergence to the ground state of the system Hamiltonian such as those discussed in~\cite{polla2021}.

The geometrical locality of Hamiltonians can also be used to improve the precision of the universal algorithm. It is possible, for example, to reduce the resources required for low-dimensional Hamiltonians using the techniques proposed in \cite{biglin2010}. Furthermore, the error can be reduced by using higher-order product formulas \cite{childs2019} and randomized gate positions \cite{campbell2019random}. For both algorithms, error mitigation can be improved by an optimization procedure that runs conditionally from layer to layer~\cite{skolik2021layerwise}, as well as, other techniques in machine learning and in particular supervised machine learning.\\

 \textbf{Available Code}. The code used to generate presented data is available at:\\ \url{https://github.com/IBM/gibbs-qalgrthms/tree/latest}\\

 \textbf{Acknowledgements}. -- We would like to thank David Layden, Sarah Sheldon, Mario Motta, Jeff Cohn, Kunal Sharma, Abhinav Deshpande, and Kristan Temme for helpful discussions. We acknowledge funding from the MIT-IBM Watson AI Lab under the project {\it Machine Learning in Hilbert space}. The research was partly supported by the IBM Research Frontiers Institute.

\let\oldaddcontentsline\addcontentsline
\renewcommand{\addcontentsline}[3]{}

\let\addcontentsline\oldaddcontentsline

\clearpage
\pagebreak

\setcounter{page}{1}
\setcounter{equation}{0}
\setcounter{figure}{0}
\renewcommand{\theequation}{S.\arabic{equation}}
\renewcommand{\thetheoremS}{S\arabic{theoremS}}
\renewcommand{\thelemmaS}{S\arabic{lemmaS}}
\renewcommand{\thefigure}{S\arabic{figure}}
\renewcommand*{\thepage}{S\arabic{page}}

\algblock{Input}{EndInput}
\algnotext{EndInput}
\algblock{Output}{EndOutput}
\algnotext{EndOutput}
\newcommand{\Desc}[2]{\State\makebox[2em][l]{#1}#2}

\onecolumngrid

\begin{center}
{\Large \textbf{Supplementary Information:\\  ``Preparing thermal states on noiseless and noisy programmable quantum processors"}}\\
\vspace{0.25cm}
{Oles Shtanko$^1$ and Ramis Movassagh$^{2,3}$}\\

\vspace{0.25cm}
\textit{\small $^1$\almaden}\\
\textit{\small $^2$\cambridge}\\
\textit{\small $^3$\google}
\end{center}

\vspace{0.5cm}

\tableofcontents

\vspace{1.0cm}

\twocolumngrid

\section{Details of ergodic algorithm} 

In this section, we provide a detailed description of the ergodic algorithm and evaluate its performance.

\subsection{Detailed description}

The algorithm works as follows: First, the system qubits are initialized in a random state in the computational basis, while the ancilla is initialized in an easily prepared thermal state. Second, the system and the ancilla undergo $d$ cycles of unitary evolution, with the ancilla reset to the thermal state after each cycle. Finally, the expected output of the $k$th cycle, $\overline \rho_k \equiv \mathbb E[\rho_k]$, satisfies the recurrence relation
\be\label{eqs:cycle_process}
\overline\rho_{k+1} = \mathcal E(\overline\rho_k) 
\ee
expressed by the quantum channel
\be\label{eqs:quantum_process}
\mathcal E(\rho) := \mathbb E\,\Tr_a(U_k\rho \otimes \sigma_{\beta k} U_k^\dag),
\ee
where $U_k$ represents unitary evolution operators, $\sigma_{\beta k}$ is the thermal Gibbs state of the ancilla at $k$th cycle, $\Tr_a$ is a partial trace taken over the ancilla degrees of freedom, and the expectation is taken over random parameters of the circuit (see below for more details).

The unitary operators are given by
\be\label{eq:ham_evolution}
U_k = \exp(-i\mathcal H_k t_k),
\ee
 where $t_k$ represents random time intervals generated from a Poisson distribution with probability density function $p(t_k) = \gamma\exp(-\gamma t_k)$, where $\gamma$ is a parameter. Additionally, $\mathcal H_k$ is given by Eq.~\eqref{eq:methdo2_ham} in the main text.

We choose the Hamiltonian for the ancilla as
\be
H^k_{\rm anc} = \frac 12 \sum_{m=1}^{n_a}\omega_{km} Z_m,
\ee
where $n_a$ is the number of ancilla qubits, $Z_m$ represents the Pauli-$Z$ operator acting on the $m$th ancilla qubit, and $\omega_{km}\in[-\Omega,\Omega]$ are frequencies chosen independently from a uniform distribution and $\Omega$ sets the frequency range. The thermal Gibbs state for such a Hamiltonian is a product of single-qubit states,
\be\label{eqs:ancilla_thermal_state}
\sigma_{\beta k} = \bigotimes_{m=1}^{n_a}\Bigl[n(\beta\omega_{km})|0\>_m\<0|_m+n(-\beta\omega_{km})|1\>_m\<1|_m\Bigl],
\ee
where $n(x) = (1+\exp(x))^{-1}$ is a single-qubit thermal distribution. To prepare this state, one could start from the product of $|0\>$ states for each ancilla qubit and apply x-axis rotations to obtain the state $|\psi_k\> = R^X_k|0\dots 0\>$, where $R^X_k$ represents the unitary operator corresponding to the x-axis rotation, applied to all ancilla qubits,
\be
R^X_k(\theta) = \exp(i\theta_{k1} X_1)\otimes\dots\otimes \exp(i\theta_{kn_a} X_{n_a}),
\ee
where $\theta_{km} = \arccos \sqrt{n(\beta\omega_{km})}$ and $X_m$ is pauli-X matrix acting on the $m$-th ancilla qubit.
After the unitary transformation, one need to implement a measurement in the computational basis and discard the result. In expectation, this step converts the pure state $|\psi_k\>\<\psi_k|$ into the mixed state in Eq.~\eqref{eqs:ancilla_thermal_state}.

Finally, the expectation in Eq.~\eqref{eqs:quantum_process} is expressed as a decomposition
\be\label{eqs:expectations}
\mathbb E = \mathbb E_\omega \mathbb E_t \mathbb E_V
\ee
of independent averages over random times $t_k$ ($\mathbb E_t$), frequencies $\omega_{km}$ ($\mathbb E_\omega$), and the random local operators $V_{km}$ in Eq.~\eqref{eqs:eth} ($\mathbb E_V$). 

In the proof of our main result, as stated in the manuscript, we assume that the matrix elements of the local coupling terms satisfy the ETH ansatz:
\be\label{eqs:eth}
\<\mu|V_{km}|\nu\> = V^{km}_\mu \delta_{\mu\nu}+\sigma^{km}_{\mu\nu}R_{\mu\nu},
\ee
where $|\mu\>,|\nu\rangle$ are eigenstates corresponding to nearby energies $|E_\mu-E_\nu|\leq \delta$ of the Hamiltonian $H$ (the energy scale $\delta$ at which ETH applies is defined by the transport properties of the system \cite{dymarsky2022}), $V^{km}_\mu$ are diagonal elements, $|\sigma^{km}_{\mu\nu}| \propto 2^{-fn}$ are the amplitudes of off-diagonal elements for some $f>0$ that depend smoothly on the energy difference $|E_\mu-E_\nu|$, and $R_{\mu\nu} \in \mathcal N(0,1)$ are assumed to be independent standard gaussians. As commonly used in the literature \cite{deutch1991quantum,srednicki1994chaos}, the ETH assumption allows us to replace the system evolution by a self-averaging over the random matrices $R$, namely
\be
\overline\rho_k \simeq \mathbb E_R \overline\rho_k.
\ee
We consider the regime where $\lambda$, $\gamma$, and their ratio $\lambda^2/\gamma$ are sufficiently small. Then the steady state of the process given by Eq.\eqref{eqs:quantum_process} converges to the Gibbs state after averaging over the random parameters in the circuit. The convergence can be given in terms of the mixing time $m$ to achieve the $\epsilon$-precision of the classical Metropolis-Hastings algorithm, 
\be\label{eqs:t_form}
\begin{split}
T_{\mu\nu} = (1-\tau_\nu)\delta_{\mu\nu}+ C\frac{2\pi\sigma^2_{\mu\nu}}{1+e^{\beta\Omega_{\mu\nu}}}f(\Omega_{\mu\nu}/\Omega),
\end{split}
\ee
where $\tau_{\nu}$ are chosen such that $\sum_{\mu} T_{\mu\nu}=1$, $\Omega_{\mu\nu} = E_\mu-E_\nu$ is the difference between the system's eigenenergies, and $\sigma^2_{\mu\nu} := \mathbb E_V |V_{\mu\nu}|^2$. Here, $f(x)$ is Heaviside-like window function has the form
\be\label{eqs:window_function}
f(x) := \frac 1\pi\Bigl[\arctan\left(\frac{\Omega}{\gamma}(1-x)\right)+\arctan\left(\frac{\Omega}{\gamma}(1+x)\right)\Bigl].
\ee
For $\gamma/\Omega\ll 1$, this function is non-zero for $x\in[-1,1]$ and limits the transition between energy levels within the energy window $[-\Omega, \Omega]$. It is also an even function $f(x)=f(-x)$.

This convergence is given by Theorem~\ref{eq:ergodic_theorem2}, whose formal version writes:

\begin{theoremS}[formal of Theorem~\ref{eq:ergodic_theorem2}]\label{ergodic_theorem_detailed} For any small parameter $0<\epsilon\leq1$ for following choice of parameters
\be
\begin{split}
&d = \tilde O(m^2/\epsilon),\\
&\gamma^{-1} = \tilde O(\beta m/\epsilon), \\
&\lambda = \tilde O(\epsilon^{3/2}/\beta m^{3/2}),\\
&n_a = \tilde O(\beta m \Omega/\epsilon),
\end{split}
\ee
we have
\be
 \| \mathbb E[\rho_d]-\rho_\beta\|_1\leq\epsilon.
\ee
Using these scaling, the implementation physical time of the algorithm is $t = \tilde O(\beta m^3/\epsilon^2)$.
\label{thm:2s}
\end{theoremS}

The proof of the theorem can be found in the following section. As part of the proof, we will show that for any state of the form $\rho = \sum_\mu p_\mu|\mu\>\<\mu|$, there exists a proper choice of $\lambda$, $\gamma$, and $n_a$ such that the process described by Eq.~\eqref{eqs:quantum_process} satisfies
\be\label{eqs:markov_update_rule}
\begin{split}
\mathcal E(\rho) \approx (1-\alpha)\sum_\mu p_{\mu}|\mu\>\<\mu|+&\alpha\sum_{\mu\nu}  T_{\mu\nu}p_{\nu}|\mu\>\<\mu|,
\end{split}
\ee
where the transition amplitudes $T_{\mu\nu}$ correspond to a classical Metropolis-Hastings algorithm in Eq.~\eqref{eqs:t_form} and 
\be\label{eqs:D-d_prop}
\alpha = \frac{n_a\lambda^2}{\gamma \Omega C}\ll1
\ee
is a small parameter. Note that $C$ is a constant with an arbitrary positive value, since it does not affect the map in Eq.~\eqref{eqs:markov_update_rule}. However, it is crucial to choose $C$ such that $\tau_\nu\geq0$ ensures that $T_{\mu\nu}$ can be interpreted as the transition matrix for a stochastic process. 

Using Eq.~\eqref{eqs:markov_update_rule}, we will show that for sufficiently large values of $d$, the expected output of the circuit is well approximated by $\overline \rho_d \simeq \sum_\mu p^{(r)}_{\mu}|\mu\>\<\mu|$, where the probabilities $p^{(r)}_{\mu}$ are generated by $r\propto \alpha d$ Metropolis update steps defined by the matrix $T$ in Eq. ~\eqref{eqs:t_form}. Any transition matrix $T$ with strictly positive entries is known to be a contractive map~\cite{levin2017markov} characterized by a non-zero gap $\Delta$. For any such map, the mixing time is $m\propto 1/\Delta$ with a prefactor that scales at most linearly with the number of qubits.
Since under ETH $\sigma^2_{\mu\nu}$ are positive functions of the energy difference $\Omega_{\mu\nu}$, one can deduce from Eq.~\eqref{eqs:t_form} that $T_{\mu\nu}>0$. The contractive property together with the classical detailed balance condition imply that Eq.~\eqref{eqs:markov_update_rule} leads to a unique steady state that is $\epsilon$-close to the Gibbs state. 

We evaluate the gap for $T$ in Eq.~\eqref{eqs:t_form} for a size-limited numerical simulation of the hardcode Bose-Hubbard Hamiltonian in Eq.~\eqref{eq:hamiltonian} as shown in Fig.~\ref{fig:fig1} of the main text. We focus only on the reduced space including $n/2$ bosons and as $V_{km}$ we use the density operators $n_i = a_i^\dag a_i$ and obtain
\be
T_{\mu\nu} = (1-\tau_\nu)\delta_{\mu\nu}+C\frac{f(\Omega_{\mu\nu}/\Omega)}{1+e^{\beta\Omega_{\mu\nu}}}\frac{1}{n}\sum_{i=1}^n|\<\mu|n_i|\nu\>|^2,
\ee
where we have chosen $C$ so that $\max(\sum_{\mu}\tau_{\mu\nu})=1$.
This simulation suggests that the gap has a reasonable polynomial scaling with system size.

\subsection{Proof of Theorem \ref{ergodic_theorem_detailed}}

As a first step, it is convenient to represent the unitary transformation in Eq.~\eqref{eq:ham_evolution} using the Liouville superoperator $\mathcal L_k(\rho) = -i[\mathcal H_k,\rho]$. Then we can rewrite
\be
 U_k \rho U^\dag_k = \exp(\mathcal L_k t_k)( \rho).
\ee
Next, we use the decomposition $\mathcal L_k = \mathcal L^0_{k} +\lambda \mathcal L^1_k$, following the structure of the Hamiltonian in Eq.~\eqref{eq:methdo2_ham} in the main text. Here,
\be
\begin{split}
&\mathcal L^0_k(\rho) = -i\Bigl[H\otimes I_{\rm anc}+I_{\rm sys}\otimes H^k_{\rm anc},\;\rho\;\Bigl],\\
&\mathcal L^1_k(\rho) = -i\sum_{m=1}^{n_a}[V_{km}\otimes X_m,\rho],
\end{split}
\ee
and $I_{\rm sys}$, $I_{\rm anc}$ are the identity operators on the spaces of system and ancilla qubits, respectively.

The main channel (Eq.\eqref{eqs:quantum_process}) in the superoperator representation includes expectation given by Eq.~\eqref{eqs:expectations} and can be written as
\be\label{eqs:laplace}
\begin{split}
\mathcal E(\rho) &= \mathbb E_V \mathbb E_\omega \mathbb E_t\Tr_a \Biggl[\exp(\mathcal L_k t) \left(\rho\otimes\sigma_{\beta k}\right)\Biggl] \\
&= \gamma\mathbb E_V \mathbb E_\omega \int_0^\infty dt e^{-\gamma t}\Tr_a \Biggl[\exp(\mathcal L_k t) \left(\rho\otimes\sigma_{\beta k}\right)\Biggl]  \\
&=\gamma\mathbb E_V \mathbb E_\omega\Tr_a \Biggl[\frac1{\gamma-\mathcal L^0_k-\lambda\mathcal L^1_k}\left(\rho\otimes\sigma_{\beta k}\right)\Biggl],
\end{split}
\ee
where we used the Laplace transform of the exponential function to express the cycle time average.

The next step is to express the propagator in Eq.~\eqref{eqs:laplace} as a Dyson's series in terms of the parameter $\lambda$ \cite{economou2006green}, resulting in
\be\label{eqs:dyson}
\begin{split}
\frac1{\gamma-\mathcal L^0_k-\lambda\mathcal L^1_k} =\frac1{\gamma-\mathcal L^0_k} \sum_{\ell=0}^\infty\lambda^\ell\Bigl(\mathcal L^1_k\frac1{\gamma-\mathcal L^0_k}\Bigl)^\ell.
\end{split}
\ee
When we insert this series into Eq.~\eqref{eqs:laplace}, we observe that all contributions from odd $\ell \in 2\mathbb N+1$ vanish exactly. This result stems from the fact that the ancilla is initialized in the computational basis and the superoperator $\mathcal L^1_k$ is proportional to the  Pauli-$X$ operator acting on either the left or right side of the density operator of a single ancilla qubit. As a result, odd numbers of $\mathcal L^1_k$ always produce a matrix with zero diagonals, which then vanish upon partial trace.

Then, the expression in Eq.~\eqref{eqs:laplace} can be written as
\be\label{eqs:2-4_representation}
\begin{split}
\mathcal E(\rho) = C_0(\rho) +\lambda^2C_2(\rho) + \lambda^4 C_4(\rho),
\end{split}
\ee
where $C_0(\cdot)$, $C_2(\cdot)$, and $C_4(\cdot)$ are linear maps defined as
\be\label{eqs:maps}
\begin{split}
&C_0(\rho) := \gamma\frac1{\gamma-\mathcal L^0_k}(\rho),\\
&C_2(\rho) := \gamma\,  \mathbb E_V\mathbb E_{\omega} \Tr_a\left[ \Bigl(\frac1{\gamma-\mathcal L^0_k}\mathcal L^1_k\Bigl)^2\frac1{\gamma-\mathcal L^0_k} \left(\rho\otimes \sigma_{\beta k}\right)\right],
\end{split}
\ee
and
\be\label{eqs:maps2}
C_4(\rho) := \gamma \mathbb E_{V}\mathbb E_\omega\Tr_a\left[ \Bigl(\frac1{\gamma-\mathcal L^0_k}\mathcal L^1_k\Bigl)^4 \frac1{\gamma-\mathcal L_k}\left(\rho\otimes \sigma_{\beta k}\right)\right].
\ee
In the last expression, the rightmost propagator contains the full Liouvillian operator $\mathcal L_k$, accounting for the higher orders in the perturbation series, making the equality in Eq.~\eqref{eqs:2-4_representation} exact.

We now analyze the action of the perturbation $\mathcal L^1_k$ and the unperturbed propagator $\mathcal L^0_k$ on a basis state. First, the action of the perturbation is
\be
\begin{split}
\mathcal L^1_k\Bigl( |\mu\>\<\nu|&\otimes |\vec a\>\<\vec b|\Bigl) \\
= & -i \sum_{\delta,m} V^{km}_{\delta\mu}|\delta\>\<\nu|\otimes |..1-a_m..\>\<..b_m..|\\
&+i\sum_{\delta,m} V^{km}_{\nu\delta}|\mu\>\<\delta|\otimes |..a_m..\>\<..1- b_m..|,
\end{split}
\ee
where $|\mu\>$ are the eigenstates of the target Hamiltonian $H$ and $|\vec a\> = |a_1\dots a_{n_a}\>$ are the eigenstates of the ancilla Hamiltonian, which are product states in the computational basis. We also used the notation $V^{km}_{\mu\nu}:=\<\mu|V_{km}|\nu\>$. Also, we use the expression
\be
\frac{1}{\gamma-\mathcal L^0_k}  |\mu\>\<\nu|\otimes |\vec a\>\<\vec b| = \frac1{\gamma+i(\Omega_{\mu\nu}+\omega_{\vec a\vec b})}|\mu\>\<\nu|\otimes |\vec a\>\<\vec b|,
\ee
where $\Omega_{\mu\nu} = E_\mu-E_\nu$, $\omega_{\vec a\vec b} = e_{\vec a}-e_{\vec b}$, and $e_{\vec a}$ are the eigenenergies of the ancilla Hamiltonian.

The action of the second maps is
\be\label{eqs:A_expression}
\begin{split}
 C_2(|\nu\>\<\nu'|) :=\; &\gamma\mathbb E_\omega\mathbb E_V\sum_{m=1}^{n_a}\sum_{s=\pm1} \frac{n(s\omega_{km})}{\gamma+i\Omega_{\nu\nu'}}\times\\
&\sum_{\alpha\mu}\Biggl( \frac{V^{km}_{\mu\nu}V^{km}_{\nu'\alpha}|\mu\>\<\alpha|}{(\gamma + i(\Omega_{\mu\nu'}-s\omega_{km}))(\gamma+i\Omega_{\mu\alpha})}\\
&\quad+ \frac{V^{km}_{\nu'\mu}V^{km}_{\alpha\nu}|\alpha\>\<\mu|}{(\gamma + i(\Omega_{\nu\mu}+s\omega_{km}))(\gamma+i\Omega_{\alpha\mu})}\\
&\quad -\frac{V^{km}_{\mu\nu}V^{km}_{\alpha\mu}|\alpha\>\<\nu'|}{(\gamma + i(\Omega_{\mu\nu'}-s\omega_{km}))(\gamma+i\Omega_{\alpha\nu'})}\\
&\quad -\frac{V^{km}_{\nu'\mu}V^{km}_{\mu\alpha}|\nu\>\<\alpha|}{(\gamma + i(\Omega_{\nu\mu}+s\omega_{km}))(\gamma+i\Omega_{\nu\alpha})}\Biggl).
\end{split}
\ee

Next, we use the ETH assumption to ensure that
\be\label{eqs:corr}
C_2(\rho) \simeq \mathbb E_R C_2(\rho).
\ee
In particular, this assumption implies that
\be\label{eqs:matrix_elements_corrf}
\begin{split}
 \mathbb E_R  V^{km}_{\mu\nu}V^{km}_{\nu'\mu'} =  (V_\mu^{km})^2\delta_{\mu\nu}\delta_{\mu'\nu'}+(\sigma^{km}_{\mu\nu})^2\delta_{\mu\mu'}\delta_{\nu\nu'}.
 \end{split}
 \ee 

Let us consider the contributions for the diagonal and off-diagonal entries only,
\be\label{eqs:expectationR_of_A}
\begin{split}
C_2(|\nu\>\<\nu|) =\; &2\mathbb E_\omega\sum_{\mu}\sum_{s=\pm1}\sum_{m=1}^{n_a}\sigma_{\mu\nu}^2\times\\
 & \frac{n(s\omega_{km})}{(s\omega_{km}-\Omega_{\mu\nu})^2+\gamma^2}\Bigl(|\mu\>\<\mu|-|\nu\>\<\nu|\Bigl),
\end{split}
\ee
where we defined
\be
\sigma^2_{\mu\nu} := \mathbb E_V \Bigl\{(\sigma^{km}_{\mu\nu})^2+(V^{km}_{\mu})^2\delta_{\mu\nu}\Bigl\} \simeq \mathbb E_V |V_{\mu\nu}|^2.
\ee
Note that this transformation does not depend on the diagonal terms. The next step would be to calculate the expectation $\mathbb E_\omega$ over the frequencies of the ancilla qubits, which can be done with the formula
\be\label{eqs:freq_average}
\mathbb E_\omega := \frac{1}{(2\Omega)^{n_a}}\int_{-\Omega}^\Omega d\omega_{k1}\dots \int_{-\Omega}^\Omega d\omega_{kn_a }.
\ee
The corresponding integrals can subsequently be computed by employing the subsequent result:

\begin{lemmaS} \label{lemma_int} Assume $\beta\gamma\ll1$. Then
\be
\begin{split}
\int_{-\Omega}^\Omega   \frac{ d\omega n(\beta\omega)}{(\omega-\Omega_{\mu\nu})^2+\gamma^2} = \frac{\pi}{\gamma} \Bigl(n(\beta \Omega_{\mu\nu})f(\Omega_{\mu\nu}/\Omega)+O(\beta\gamma)\Bigl),
\end{split}
\ee
where $f(x)$ is defined in Eq.~\eqref{eqs:window_function}.
\end{lemmaS}

In Section \ref{proof_lemma_int}, we provide a proof of this Lemma. With the help of its results, we can transform Eq.~\eqref{eqs:expectationR_of_A} into 
\be\label{eqs:c2_integral_expression}
\begin{split}
\lambda^2 C_2\bigl(|\nu\>\<\nu|\bigl) =\frac {2\pi n_a\lambda^2}{\Omega\gamma} &\sum_{\mu} \sigma^2_{\mu\nu}n(\Omega_{\mu\nu})f(\Omega_{\mu\nu}/\Omega)\\&\times \Bigl(|\mu\>\<\mu|-|\nu\>\<\nu|\Bigl)+O\Bigl(\frac {\beta \lambda^2 n_a}{\Omega}\Bigl).
\end{split}
\ee
where, here and below, we use $O$-notation for quantum states and operators in a sense of the \textit{trace norm} of the remaining terms. 

As we will see below, the last $O$-small term constitute a part of algorithm's error. Next, we choose main parameters to be functions of $\epsilon'$ with scaling
\be\label{eqs:scaling_formulas1}
\begin{split}
n_a = O(\beta\Omega/\epsilon'),\quad \gamma^{-1} = O(\beta/\epsilon'), \quad \lambda = O(\epsilon'^{3/2}/\beta).
\end{split}
\ee
This choice allows us to rewrite the scaling of the error terms in Eq.~\eqref{eqs:c2_integral_expression} as
\be\label{eq:eqs:scaling}
\begin{split}
\frac{\lambda^2n_a}{\gamma\Omega} =  O(\epsilon'), \qquad \frac {\beta \lambda^2n_a}{\Omega}= O(\epsilon'^2)
\end{split}
\ee
  
  Let us assume now that the expected density matrix $\overline\rho_k$ is nearly diagonal in the energy basis, i.e.
\be\label{eqs:ansatz}
\overline \rho_k = \tilde \rho_k+\delta\rho_k, \qquad \tilde \rho_k = \sum_\mu p_{k\mu}|\mu\>\<\mu|.
\ee
where $p_{k\mu}$ are non-negative occupations of the energy levels and $\delta\rho_k$ are small corrections that satisfy $\|\delta\rho_k\|_1 \le \epsilon_k$. From Eqs.~\eqref{eqs:c2_integral_expression} and \eqref{eq:eqs:scaling}, it follows that the diagonal part undergoes transformation as
\be\label{eqs:2nd_order_o_expression}
\begin{split}
\lambda^2 C_2(\tilde\rho_k) = \alpha\sum_{\mu\nu} \tau_{\mu\nu}p_{k\nu}\Bigl(&|\mu\>\<\mu|-|\nu\>\<\nu|\Bigl)+O(\epsilon'^2).
\end{split}
\ee
where we used the definition of $\alpha$ in Eq.~\eqref{eqs:D-d_prop} and amplitudes $\tau_{\mu\nu}$ in Eq.~\eqref{eqs:t_form}. Note that $\alpha = O(\epsilon')$, the property that we will use later.

At the same time, the effect of $C_2$ on the non-diagonal correction can be characterized by the following Lemma.

\begin{lemmaS} \label{c2_lemma} Given the scaling of the parameters $\gamma^{-1}$, $\lambda$, and $n_a$ in Eq.~\eqref{eq:eqs:scaling}, for any $\delta\rho$, it follows that
\be
\begin{split}
\lambda^2\| C_2(\delta\rho)\|_1 = \tilde O\left(\epsilon'\|\delta\rho\|_1\right).
\end{split}
\ee
\end{lemmaS}

The proof of this Lemma can be found in Section.~\ref{proof_lemma_c2} below. Combining these two expressions, we get
\be\label{eqs:2nd_order_o_expression2}
\begin{split}
\lambda^2 C_2(\overline\rho_k) = \alpha\sum_{\mu\nu} \tau_{\mu\nu}p_{k\nu}\Bigl(&|\mu\>\<\mu|-|\nu\>\<\nu|\Bigl)\\
&+\tilde O(\epsilon'\epsilon_k)+\tilde O(\epsilon'^2).
\end{split}
\ee
  Next, we utilize the following Lemma to establish the bound for the operator $C_4(\rho)$.

\begin{lemmaS} \label{c4_lemma} Given the scaling of the parameters in Eq.~\eqref{eq:eqs:scaling}, for any density operator $\rho$, it follows that
\be\label{eqs:C4_condition}
\lambda^4\|C_4(\rho) \|_1 = \tilde O(\epsilon'^2).
\ee
\end{lemmaS}
The proof can be found in Sec.~E below. Using the recurrence relation in Eq.~\eqref{eqs:cycle_process}, we derive that
\be
\begin{split}
\overline \rho_{k+1} = \tilde \rho_{k}+C_0(\delta\rho_k)+\alpha\sum_{\mu\nu} \tau_{\mu\nu}p_{k\nu}\Bigl(&|\mu\>\<\mu|-|\nu\>\<\nu|\Bigl)\\
&+\tilde O(\alpha \epsilon_k)+\tilde O(\epsilon'^2),
\end{split}
\ee
In terms of the diagonal part and the magnitude of the correction in Eq.~\eqref{eqs:ansatz}, this relation van be written in the form
\be
\begin{split}
&\tilde \rho_{k+1}  = \mathcal V(\tilde \rho_k),\\
&\epsilon_{k+1} = \epsilon_k + O(\epsilon'\epsilon_k)+\tilde O(\epsilon'^2),
\end{split}
\ee
where $\mathcal V = (1-\alpha)\mathcal I+\alpha \mathcal T$, and action of the map $\mathcal T$ on the diagonal matrix is defined as
\be
\mathcal T(\tilde \rho_k) = \sum_\nu T_{\mu\nu}p_{k\nu} |\mu\>\<\mu|,
\ee
where $T_{\mu\nu}$ is defined in Eq.~\eqref{eqs:t_form}.
By iterating this map while assuming $\epsilon_0 = 0$, we arrive at
\be
\overline\rho_k = \mathcal V^k(\rho_0)+\delta\rho_k, \qquad \|\delta\rho_k\|_1 = \tilde O(k\epsilon'^2).
\ee
The first term is represented by the map
\be
\mathcal V^k = \Bigl((1-\alpha)\mathcal I+\alpha\mathcal T\Bigl)^k = \sum_{l=0}^k p(l;k,\alpha)\mathcal T^l,
\ee
where $\mathcal I$ represents the identity superoperator and $p(l;k,\alpha)$ are probabilities from the binomial distribution,
\be
p(l;k,\alpha) := \binom{k}{l}\alpha^l(1-\alpha)^{k-l}.
\ee
By using the triangle inequality, we bound the difference between the main part of the Gibbs state $\rho_\beta$ as
\be\label{eqs:distance}
\begin{split}
\|\mathcal V^k(\rho_0)-\rho_\beta\|_1 &\leq\Bigl\|\sum_{l=0}^k p(l;k,\alpha) (\mathcal T^l(\rho_0)-\rho_\beta)\Bigl\|_1\\
&\leq\sum_{l=0}^k p(l;k,\alpha)\Bigl\| \mathcal T^l(\rho_0)-\rho_\beta\Bigl\|_1
\end{split}
\ee
Next, we use the inequality \cite{temme2010chi}
\be
\Bigl\| \mathcal T^l(\rho_0)-\rho_\beta\Bigl\|_1\leq q_n (1-\Delta)^{\ell}
\ee
where $\Delta$ is the gap of the process $\mathcal T$ and $q_n$ is a constant that grows exponentially with the number of qubits $n$. This leads us to
\be
\|\mathcal V^k(\rho_0)-\rho_\beta\|_1 \leq q_n\sum_{l=0}^k p(l;k,\alpha)(1-\Delta)^{\ell}.
\ee
By choosing certain integer $0<s<k$, we divide the sum into two parts as
\be
\begin{split}
\|\mathcal V^k(\rho_0)-\rho_\beta\|_1 \leq q_n\sum_{l=0}^s &p(l;k,\alpha)(1-\Delta)^{\ell}\\
&+q_n\sum_{l=s+1}^k p(l;k,\alpha)(1-\Delta)^{\ell}.
\end{split}
\ee
Each sum can be bounded by taking the smallest value in the exponent of $(1-\Delta)$, resulting in
\be
\|\mathcal V^k(\rho_0)-\rho_\beta\|_1 \leq q_n\Bigl(\sum_{l=0}^s p(l;k,\alpha)+(1-\Delta)^{\ell}\Bigl).
\ee
For all $s \le \alpha k$, we can utilize the Chernoff bound for the binomial distribution \cite{arratia1989tutorial} in the form
\be
\sum_{l=0}^s p(l;k,\alpha) \leq \exp\Bigl(-k D\left(\frac{s}{k}\|\alpha\right)\Bigl),
\ee
where $D(p\|p')$ represents the Kullback-Leibler divergence between two biased binary distributions, characterized by probabilities $p$ and $p'$, i.e.
\be
D(p\|p') =  p\log \frac{p}{p'}+(1-p)\log \frac{1-p}{1-p'}.
\ee
Now, let us choose the division parameter to be $s = \mu \alpha k$, where $0 < \mu \le 1$ is a constant. Then
\be
D\left(\frac{s}{k}\|\alpha\right) = D(\mu\alpha\|\alpha)\geq f(\mu)\alpha,
\ee
where $f(\mu) := 1-\mu + \mu\log\mu$ is a non-negative function for $\mu \in (0,1]$. By utilizing this property, we can state that
\be
\sum_{l=0}^{\mu\alpha k} p(l;k,\alpha) \leq \exp\left(-f(\mu)\alpha k\right),
\ee
and therefore, the distance in Eq.~\eqref{eqs:distance} satisfies
\be
\begin{split}
\|\mathcal V^k(\rho_0)-\rho_\beta\|_1 \leq q_ne^{-O(\alpha k\Delta)}.
\end{split}
\ee
 We will now recall the definition of $\alpha$ from Eq.~\eqref{eqs:D-d_prop} and use the scaling in Eq.~\eqref{eq:eqs:scaling} to show that $\alpha = O(\epsilon')$. This implies that the distance between the output of the algorithm at step $d$ and the Gibbs state is bounded by 
\be\label{eqs:diffc}
\|\rho_d- \rho_\beta\|_1 \leq  q_ne^{-O(\Delta\epsilon' d)}+\tilde O(d \epsilon'^2) :=\epsilon,
\ee
where $\epsilon$ is our target error. To compensate for the exponentially large $q_n$ factor, we need certain $d = O(m\epsilon')$, where $m = \tilde O(\log q_n/\Delta)$ is mixing time. Another consition is that the contribution from the second term is small, $\epsilon = \tilde O(\epsilon'^2 d)$. These conditions can be satisfied together if
\be
d = \tilde O(m^2/\epsilon),\qquad \epsilon' = \tilde O(\epsilon/m).
\ee
Substituting this error measure into Eq.~\eqref{eqs:scaling_formulas1}, we obtain the required scaling of the parameters
\be
\begin{split}
\gamma^{-1} = \tilde O(\beta m/\epsilon ), \quad \lambda = \tilde O(\epsilon^{3/2}/\beta m^{3/2}),
\end{split}
\ee
Finally, the average time of the algorithm must be scaled as $t := d/\gamma = O(\beta m^3/\epsilon^2)$. This completes our proof.

\subsection{Proof of Lemma~\ref{lemma_int}}
\label{proof_lemma_int}

We first rewrite the integral as
\be
\begin{split}
\int_{-\Omega}^\Omega d\omega & \frac{n(\beta\omega)}{(\omega-\Omega_{\mu\nu})^2+\gamma^2} = \int_{-\Omega}^\Omega d\omega\frac{n(\beta\Omega_{\mu\nu})}{(\omega-\Omega_{\mu\nu})^2+\gamma^2}\\
&+\int_{-\Omega}^\Omega d\omega \frac{n(\beta\omega)-n(\beta\Omega_{\mu\nu})}{(\omega-\Omega_{\mu\nu})^2+\gamma^2} \\
&\quad = \frac{\pi}{\gamma} \Bigl(n(\Omega_{\mu\nu})f(\Omega_{\mu\nu}/\Omega)+\beta \gamma  F(\beta \Omega_{\mu\nu},\beta\gamma,\beta\Omega)\Bigl),
\end{split}
\ee
we have used the notation
\be\label{eqs:f_integral}
F(z_0,g,W) := \int_{-W}^W dz\frac{n(z+z_0)-n(z_0)}{z^2+g^2}.
\ee
It is straightforward to show that
\be
\int_{-\Omega}^\Omega d\omega\frac{1}{(\omega-\Omega_{\mu\nu})^2+\gamma^2} = \frac{\pi}{\gamma }f(\Omega/\Omega_{\mu\nu}),
\ee
where $f(x)$ is defined in Eq.~\eqref{eqs:window_function}.

Our next objective is to demonstrate that the integral in Eq.~\eqref{eqs:f_integral} is bounded by a constant independent of $z_0$, $W$, or $g$. Since the denominator is symmetric with respect to zero, we can rewrite the integral as
\be
F(z_0,g,W) = \int_{-W}^{W}\frac{f(z,z_0)}{z^2+g^2}dz,
\ee
where we have denoted the symmetrized numerator as 
\be
\begin{split}
f(z,z_0) : &= \frac{1}{2}\Bigl(n(z+z_0)+n(-z+z_0)-2n(z_0)\Bigl)\\
& = \frac{1}{2}\frac{\tanh(z_0/2)(\cosh z-1)}{2\cosh^2(z_0/2)+\cosh z-1}.
\end{split}
\ee
The absolute value of the integral obeys the bound
\be
|F(z_0,g,W)| \leq \frac 12|\tanh (z_0/2)|\int_{\infty}^{-\infty} \frac{z^2dz}{(z^2+g^2)(z^2+A^2(z))},
\ee
where $A^2(z) = 2z^2\cosh^2(z_0/2)/(\cosh z-1)$ and we extended the limits of integration as the integrand is non-negative. Utilizing the facts that $z^2 + g^2 \geq z^2$ and $|\tanh(z_0/2)| \leq 1$, we obtain
\be
|F(z_0,g,W)| \leq \frac 12 \int_{-\infty}^{\infty} \frac{dz}{z^2+A^2(z)}. 
\ee
Since $A(z)$ monotonically decreases with $z$, for any $z'>0$, we can rewrite
\be
\begin{split}
|F(z_0,g,W)|& \leq \int_{-\infty}^{z'} \frac{dz}{z^2+A^2(z)}+\int_{z'}^{\infty} \frac{dz}{z^2+A^2(z)}
\\
& < \int_{-\infty}^{z'} \frac{dz}{z^2+A^2(z')}+\int_{z'}^{\infty} \frac{dz}{z^2} \\
&= \frac{\arctan(z'/A(z'))}{A(z')}+\frac{1}{z'}.
\end{split}
\ee
Since $\arctan(x)\leq \pi/2$ for all $x$  and $A(z') \geq \sqrt{2z'^2/(\cosh z' - 1)}$, we write
\be
|F(z_0,g,W)| < \frac{\pi}{2}\sqrt{\frac{\cosh z'-1}{2z'^2}}+\frac{1}{z'}.
\ee
For any $0 < z' < \infty$, the right-hand side is a constant that attains its minimum value of approximately 1.4045. This bound, however, is not particularly tight, as a quick numerical check suggests that $|F(z_0, g)| \lesssim 0.441$. 
Nevertheless, this result enables us to establish that
\be
\begin{split}
\int_{-\Omega}^\Omega  & \frac{d\omega n(\beta\omega)}{(\omega-\Omega_{\mu\nu})^2+\gamma^2} = \frac{\pi}{\gamma} \Bigl(n(\Omega_{\mu\nu})f(\Omega_{\mu\nu}/\Omega)+O(\beta\gamma)\Bigl).
\end{split}
\ee
This concludes the proof.

\subsection{Proof of Lemma~\ref{c2_lemma}}
\label{proof_lemma_c2}

We can rewrite the trace norm using the definition given in Eq.~\eqref{eqs:maps} as
\be
\begin{split}
\|C_2(\delta\rho)\|_1 &= \gamma \|\mathbb E_\omega \mathbb E_V \Tr_a (\mathcal G_k^0 \mathcal L_k^1)^2\mathcal G_k^0 (\delta\rho\otimes \sigma_{\beta k})\|_1
\end{split}
\ee
Using the superoperator structure, we get
\be
\begin{split}
\|C_2(\delta\rho)\|_1 \leq \gamma \sum_m \sum_{s,s'=\pm1}\mathbb E_\omega \mathbb E_V \|\tilde{\mathcal G}^k_0(0)& \mathcal V_{km}^s \tilde{\mathcal G}^k_0(s\omega_{km})\\
&\mathcal V_{km}^{s'}\tilde{\mathcal G}^k_0(0) \delta\rho\|_1
\end{split}
\ee
where we used the following notation for the ``displaced'' Green's function,
\be\label{eqs:displaced_superop}
\tilde{\mathcal G}^k_0(\omega) := \frac{1}{\gamma-\mathcal L_k^0-i\omega},
\ee
and the notation of the left and right actions of the coupling operator,
\be
\mathcal V_{km}^{+} := V_{km} \otimes I,\qquad \mathcal V_{km}^{-} := I\otimes V_{km}.
\ee
Next, we can upper bound the action of the operator $C_2$ on the hermitian operator $\delta\rho$ as
\be
\begin{split}
\|C_2(\delta\rho)\|_1 &\leq 4\gamma \sum_m \sum_{s=\pm1}\mathbb E_V \sigma^2_{\rm max}[\tilde{\mathcal G}^k_0(0)]\sigma_{\rm max}[\mathbb E_\omega \tilde{\mathcal G}^k_0(s\omega_{km})]\\
&\qquad\qquad\qquad\qquad\qquad\qquad\times \|V_{km}\|^2\|\delta\rho\|_1
\end{split}
\ee
where $\sigma_{\max}(\mathcal A)$ denotes the largest singular eigenvalue of the superoperator $\mathcal A$. Using the explicit form of the suoperoperator in Eq.~\eqref{eqs:displaced_superop}, we can derive that
\be\label{eqs:sigma_max}
\begin{split}
&\sigma_{\rm max}[\tilde{\mathcal G}^k_0(0)]\leq \frac{1}{\gamma}, \\
&\sigma_{\rm max}[\mathbb E_\omega\tilde{\mathcal G}^k_0(\omega)]=\frac{1}{2\Omega}\max_{\mu,\nu}\Biggl|\int_{-\Omega}^\Omega \frac{d\omega}{\gamma-i(\Omega_{\mu\nu}+\omega)}\Biggl|\\
& \qquad\qquad\qquad \quad= O\left(\frac{1}{\Omega}\log \frac{\Omega}{\gamma}\right).
\end{split}
\ee
Using this scaling, we arrive at the expression
\be
\|C_2(\delta\rho)\|_1 = O\left(\frac{\lambda^2n_a}{\gamma\Omega} \log\frac{\Omega}{\gamma}\|\delta\rho\|_1\right) = \tilde O(\epsilon'\|\delta\rho\|_1).
\ee
To make the last step, we used the proper scaling of the parameters from Eq.~\eqref{eq:eqs:scaling}. This expression conlcudes our proof.

\subsection{Proof of Lemma~\ref{c4_lemma}}
\label{proof_lemma_4}

 Using this density matrix, we can rewrite the action of the channel as $C_4(\rho) = \sum_{b=1}^\infty C^{(b)}_4(\rho)$, where
\be
C^{(b)}_4(\rho) = \gamma \mathbb E_{V}\mathbb E_\omega\Tr_a\left[ \Bigl(\mathcal G_0^k \mathcal L^1_k\Bigl)^{2+2b} \mathcal G_k^0(\rho\otimes\sigma_{\beta k})\right].
\ee
where $\mathcal G_k^0 := 1/(\gamma-\mathcal L^0_k)$.

Next we decompose the coupling superoperator as a sum over individual couplings to each ancilla qubit, i.e.
\be
\mathcal L^1_{k} = \sum_{m=1}^{n_a} \mathcal L^1_{mk},
\ee
which allows us to rewrite the channel using two components
\be
\lambda^4 C^{(1)}_4(\rho) = S_1(\rho)+S_2(\rho).
\ee
Each component corresponds to a different number of distinct operators $\mathcal L^1_{km}$. For example, the first term has the form
\be
\begin{split}
S_1(\rho) = \gamma\mathbb E_{V}\frac{\lambda^4}{2\Omega}\sum_{m=1}^{n_a}\int_{-\Omega}^\Omega d\omega_{km}\Tr_a\Bigl[ \Bigl(&\mathcal G^0_k\mathcal L^1_{km}\Bigl)^4 \mathcal G_k^0(\rho\otimes\sigma_{\beta k})\Bigl].
\end{split}
\ee
 The expression on the right has one integral over frequencies $\omega_{km}$ and four poles in the form $1/(\omega_{km}\pm i\gamma+\dots)$ for each operator $\mathcal G_k^0$. Taking the integral eliminates one of the poles leading to the expression in Eq.~\eqref{eqs:sigma_max}, while the remaining poles generate at most $O(\gamma^{-3})$ divergence in the limit $\gamma\to\infty$. As a result, this term behaves as 
\be
\|S_1(\rho)\|_1 = O\left(\frac{\lambda^4n_a}{\gamma^3\Omega}\log\frac{\Omega}{\gamma}\right) = \tilde O(\epsilon'^2),
\ee
where the factor of $n_a$ is due to the sum taken over ancilla qubits.

Next, the second term involves operators $\mathcal L^1_{km}$ for two distinct values of $m$, and can be written as
\be
\begin{split}
S_2(\rho) = \gamma\mathbb E_{V}\frac{\lambda^4}{4\Omega^2}&\tilde{\sum_{m_1,m_2}}\int_{-\Omega}^\Omega d\omega_{km_1}\int_{-\Omega}^\Omega d\omega_{km_2}\times \\
&\Biggl(\Tr_a\left[ \Bigl(\mathcal G^0_k\mathcal L^1_{km_1}\Bigl)^2\Bigl(\mathcal G_k^0\mathcal L^1_{km_2}\Bigl)^2 \mathcal G_k^0(\rho\otimes\sigma_{\beta k})\right]\\
&+\Tr_a\left[ \Bigl(\mathcal G^0_k\mathcal L^1_{km_1}\mathcal G^0_k\mathcal L^1_{km_2}\Bigl)^2 \mathcal G_k^0(\rho\otimes\sigma_{\beta k})\right]\Biggl),
\end{split}
\ee
where here and below the sum $\tilde{\sum}_{m_1\dots m_2}$ is taken over all values of $m_i$ such that $m_1\neq m_2$. Using the same analysis, we conclude that this expression has two pole-reducing integrals and four poles in total, and therefore has a divergence of $O(\gamma^{-2})$. The resulting bound is
\be
\|S_2(\rho)\|_1 = O\Bigl(\frac{n_a^2
\lambda^4}{\gamma^2\Omega^2}\log^2\frac{\Omega}{\gamma}\Bigl)= \tilde O(\epsilon'^2).
\ee
Application of the triangle inequality gives
\be
\lambda^4\|C^{(1)}_4(\rho)\|_1 \leq \|S_1(\rho)\|_1+\|S_2(\rho)\|_1 = \tilde O(\epsilon'^2).
\ee
In a similar fashion, the contributions of higher-order terms are
\be
\|C_4^{(b)}\|_1 = \tilde O\left(\frac{\lambda^{2+2b}}{\gamma^{2+2b}}{\rm poly}\Bigl(\frac{n_a\gamma}{\Omega}\log \frac\Omega\gamma\Bigl)\right) = \tilde O(\epsilon'^{1+b}).
\ee
Assuming that the Dyson's series converges, we get
\be
C_4(\rho) = \tilde O(\epsilon'^2).
\ee
This expression concludes the proof.

\subsection{Comparison to quantum Metropolis algorithm}\label{Supp2:comparison Metropolis}

 In this section, we compare the performance of our algorithm to that of the quantum Metropolis algorithm presented in Ref.~\cite{Temme2011}. Quantum Metropolis algorithms iteratively implement a quantum map $\mathcal E_{QM}$ that satisfies conditions similar to the detailed balance conditions of its classical counterpart. The authors provide an estimate of the distance between the fixed point $\sigma^*$ of the map $\mathcal E_{QM}$, i.e. $\mathcal E_{QM}(\sigma^*)=\sigma^*$, and the Gibbs state as
\be\label{eqs:error_scaling}
\epsilon := \|\sigma^*-\rho_\beta\|\leq \frac{\epsilon_{sg}}{1-\eta},
\ee
where $\eta$ is contraction parameter of the single-step quantum Metropolis map (i.e. for all $\rho,\sigma$ it satisfies $\|\mathcal E_{QM}(\rho-\sigma)\|_1\leq \eta\|\rho-\sigma\|_1$) and $\epsilon_{sg}$ is the error of a single quantum Metropolis step. This error has the following scaling
\be
\epsilon_{sg} = O\bigl(\beta m_{QM}/T \bigl),
\ee
where $m_{QM}$ is the mixing time of the quantum Metropolis process and $T$ is the time of implementation of a single step. The total time of the quantum Metropolis algorithm scales as
\be
t_{QM} \propto mT = O\Bigl(\frac{\beta m_{QM}^2}{(1-\eta)\epsilon}\Bigl).
\ee
We can compare this time to the runtime of our algorithm, $t = O(\beta m^3/\epsilon^2)$. Our algorithm has a more restrictive scaling with the target error compared to the quantum Metropolis algorithm, which has only a linear scaling. It is worth noting that the $1/\epsilon$ error in Eq.~\eqref{eqs:error_scaling} ignores the mixing error that we take into account. It is also difficult to compare the scaling of $m_{QM}^2/(1-\eta)$ for the quantum Metropolis algorithm with the scaling factor $m^3$ in our problem. However, it is easier to find $m$ by analyzing the classical stochastic process represented by the transition matrix $T$ in our case. Both algorithms are likely to have polynomial factors for a wide range of quantum Hamiltonians.

\subsection{Resource estimation}\label{Supp2:ResourceEstim_Erog}

Let us summarize the resources needed to implement the algorithm. As in the main text, $n$ is the number of qubits for the target Hamiltonian, $\beta$ is the inverse temperature, and $\epsilon$ is the error of the algorithm.
\begin{table}[h!]
\begin{tabular}{|l|c|}
\hline
Type of Hamiltonians & ergodic (satisfy ETH)                                   \\ \hline
Performance time     & $O(\beta n^3/\epsilon^2\Delta^3)$                          \\ \hline
Min \# of ancilla    & $O(\beta n^2/\epsilon\Delta)$                          \\ \hline
Min depth    & $O(n^{o(1)}(\beta n^3/\epsilon^2\Delta^3)^{1+o(1)})$                          \\ \hline
\end{tabular}
\end{table}

The number of gates required is for the optimal trotterized decomposition of the Hamiltonian for lattice systems evolution using product formulas \cite{childs2019}.

\subsection{Details of simulations}\label{Supp2:B}

Let us outline the details of the numerical simulations presented in Fig.2. In these simulations, we consider a one-dimensional architecture where we use a single ancilla per system qubit. We choose the local coupling operator in Eq.\eqref{eq:methdo2_ham} as $V_{km} = a_{km}X_{s(m)}+b_{km}Z_{s(m)}$, where $s(m)$ is the index of the system qubit coupled to the $m$th ancilla qubit and $a_{km},b_{km}\in\mathcal N(0,1)$ are generated from a normal distribution. We simulate the action of $d$ maps in Eq.~\eqref{eqs:quantum_process}, where the expectation is taken over parameters $\{t_k,\omega_{km}, a_{mk},b_{mk}\}$ and $U_k$ are defined in Eq.~\eqref{eq:ham_evolution}. The probability of sampling eigenstate $\mu$ and the output energy are
\be
p_\mu = \<\mu|\mathbb E[\rho_d]|\mu\>, \quad E = \Tr\bigl(H\mathbb E[\rho_d]\bigl).
\ee
In order to simulate the noisy dynamics, we consider the dynamics
\be
\rho^{\rm noisy}_{d} = \mathbb E\,\mathcal E^{\rm noisy}_{d} \circ \dots \circ \mathcal E^{\rm noisy}_{1}(\rho_0),
\ee
where, we the noisy cycles are modeled by the map
\be
\mathcal E^{\rm noisy}_k := \mathcal D_n\circ\dots \circ\mathcal D_1\circ\mathcal E_k
\ee
where $\mathcal D_m$ is single-qubit depolarizing channel acting on qubit $m$. It has the form 
\be\label{eqs:noisy_transform}
\begin{split}
\mathcal D_m(\rho) =&\bigl(1-3p\bigl)\rho + p\Bigl(X_m\rho X_m+Y_m\rho Y_m+Z_m\rho Z_m\Bigl),
\end{split}
\ee
where $p$ is the error probability, $X_m$, $Y_m$ and $Z_m$ are Pauli matrices acting on system qubit $m$.

\section{Details of universal algorithm} \label{Supp1}

In this section, we examine the universal algorithm in greater detail and analyze its convergence.

\subsection{Description and regime}

This algorithm consists of the following steps:

\begin{enumerate}
\item Initialize the system qubits in a random product state;
\item Set the ancilla qubits in the state $|0\rangle$;
\item Apply a cycle consisting of random gates to the system and ancilla qubits and measure the ancilla;
\item If at least one ancilla returns $1$, reject the output and re-run the circuit. Otherwise, repeat step 2 if the number of completed cycles is smaller than $d$;
\item Measure the output.
\end{enumerate}
    
\begin{figure}[t]
    \centering
  \includegraphics[width=\figsizenew\linewidth]{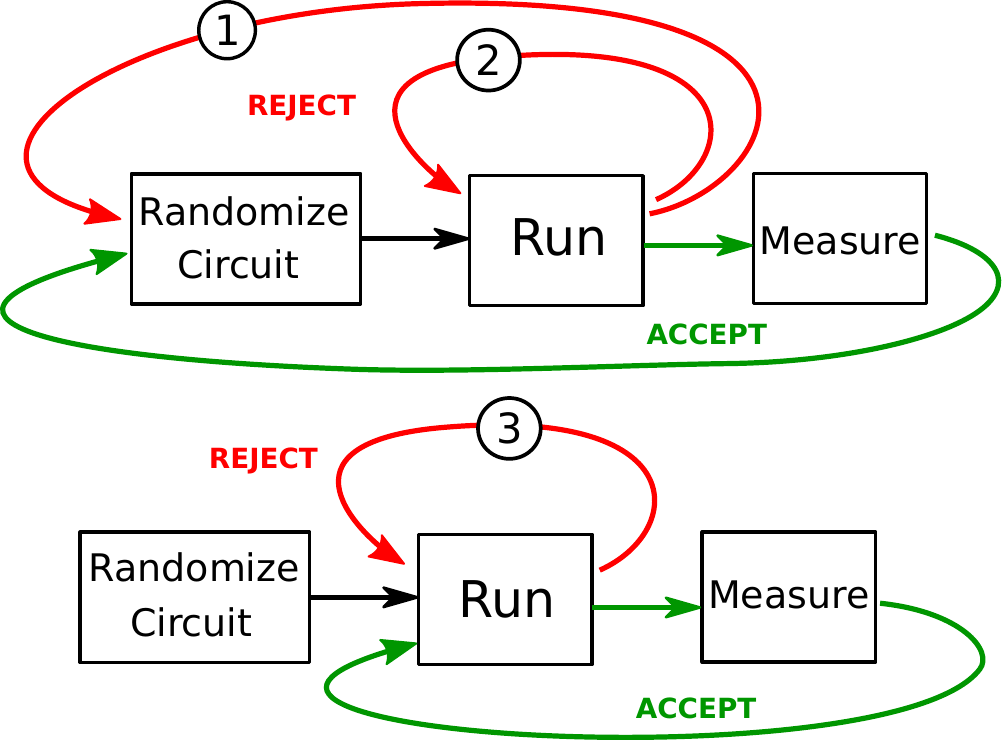}
  \caption{\textbf{Modes of operation for the universal algorithm}. In mode 1, the circuit is randomized each time it is run. In mode 2, we run the same circuit a fixed number of times and randomize only after success. In mode 3, the circuit is randomly selected at the beginning of the experiment and remains fixed for the duration of the experiment.}
  \label{fig:fig2s}
\end{figure}

There are three ways (modes) to implement the algorithm, depending on how we re-run the circuit. These modes are illustrated in Fig.~\ref{fig:fig2s}.

In the first mode, the circuit is randomized every time we run it, regardless of whether the circuit is accepted or rejected. This allows us to obtain the average outcome. The resulting distribution is described by a circuit-averaged density matrix
\be\label{eq:expect_avrg_rho}
\mathbb E[\rho_{\rm out}]= \mathbb E_\theta[P(\boldsymbol\theta)\rho_{\rm out}(\boldsymbol\theta)]/\mathbb E_\theta [P(\boldsymbol\theta)],
\ee
where $\rho_{\rm out}(\boldsymbol\theta)$ is the accepted output state and $P(\boldsymbol\theta)$ is probability of accepting the circuit, and the expectation is taken with respect to the random angles $\boldsymbol\theta = \{\theta_{km}\}$. In the second mode, the circuit is only randomized if it is accepted. In this way, we obtain the distribution of accepted outcomes described by a conditionally averaged density matrix $\rho_{\rm out} = \mathbb E_\theta \rho_{\rm out}(\boldsymbol\theta)$. In the third mode, the circuit is not randomized at all, except the first round. In this way, we obtain the distribution of outcomes for a fixed circuit. The resulting distribution is described by the density matrix $\rho_{\rm out}(\boldsymbol\theta)$.

The convergence of the algorithm is described by Theorem~\ref{thm:main_text_universal} in the main text, as well as in the extended Theorem~\ref{thm:suppl_univ} below. In particular, Theorem~\ref{thm:main_text_universal} describes the algorithm's performance for mode 1 and mode 3. The second mode is described in Theorem~\ref{thm:suppl_univ}, which also describes all three cases together.
 
This work utilizes all three modes of the algorithm. For example, the averaged simulated curves in Fig.~\ref{fig:fig2}(a,b) in the main text are computed using mode 1. The optimized curves on the same plots, however, require the use of fixed gate angles throughout the experiment, which is equivalent to mode 3. Finally, IBMQ experiments are conducted using mode 2 because it requires fewer communication rounds with the device. Below, we analyze the performance of all three modes.

\subsection{Undivided Hamiltonian ($M=1$)}

Let us first consider a non-trotterized version of this algorithm that can be applied to arbitrary, non-local Hamiltonians. This case is equivalent to the number of ancilla being $M=1$, when the Hamiltonian has only a single (global) term. In the section below, we will obtain a more general result for $M>1$. For simplicity, we will focus on mode 1 here.

We begin by initializing the system in a random state on the $n$ qubits, and the single-qubit ancilla to $|0\rangle$. To implement the algorithm, we apply a circuit containing $d$ non-local $(n+1)$-qubit unitary gates defined by
\begin{equation}\label{eq:basic_gate}
U(\theta_k,H) := \exp\Bigl(i\theta_k\sqrt{\beta H/d}\otimes X\Bigl),
\end{equation}
where $\theta_k\in \mathcal{N}(0,1)$ are independent Gaussian random variables, the Pauli operator $X$ acts on the ancilla qubit, and the Hamiltonian $H$ acts on the remaining $n$ system qubits. We omitted index $m$ for brevity as we have only one global $n$-qubit gate in this case. After each layer, we measure the ancilla qubit, reset it, and proceed to the next layer if the measurement returns zero. If the measurement returns a non-zero value, we repeat the circuit, resetting the values of $\theta_k$ before retrying.

Let us set the initial state of the system and ancilla qubit as $|\Psi_0\rangle = |\vec{z}\rangle \otimes |0\rangle$, where $|\vec{z}\rangle = |z_1, \dots, z_n\rangle$ is a random initial state of the system and $z_i$ are computational basis states, $z_i\in \{0, 1\}$. Then, the first global gate returns the state
\be
\begin{split}
U(\theta_1,H)|\Psi_0\> =& \cos\bigl(\theta_1 \sqrt{\beta H/d}\bigl)|\vec z\>\otimes|0\>\\
&+i\sin\bigl(\theta_1 \sqrt{\beta H/d}\bigl)|\vec z\>\otimes|1\>.
\end{split}
\ee

Let us define $\mathbb E[\rho_k]$ as the expected output after applying $k$ gates and accepted measurements, including averaging over the initial states. After first gate, it has the form
\be
\begin{split}
\mathbb E[\rho_1] &\propto \mathbb E_{\theta_1,\vec z} \Tr_a\Bigl\{I_{\rm sys}\otimes|0\>\<0| U(\theta_1, H)|\Psi_0\>\<\Psi_0|U^\dag(\theta_1, H)\Bigl\} \\
&= \frac 1{2^n}\mathbb E_{\theta_1}\cos^2(\theta_1 \sqrt{\beta H/d})\\
&=\frac 1{2^{n+1}}\mathbb E_{\theta_1}\Bigl(I_{\rm sys} + \cos(2\theta_1 \sqrt{\beta H/d})\Bigl)\\
&=\frac{1}{2^{n+1}}\Bigl(I_{\rm sys}+\exp(-2\beta H/d)\Bigl).
\end{split}
\ee
As a result of repeating this procedure, it is straightforward to show that $d$-cycle algorithm returns the expected state
\be\label{eqs:r_bd}
\begin{split}
\mathbb E[\rho_{\rm out}] \equiv \mathbb E[\rho_d] & \propto \frac 1{2^{n+d}}\Bigl(I+\exp(-2\beta H/d)\Bigl)^d.
\end{split}
\ee
Upon normalization, and using the binomial theorem, we obtain
\be\label{eqs:binom_gibbs}
\mathbb E[\rho_{\rm out}]  =  \frac 1{\mathcal Z_{\beta,d}}\sum_{k=0}^d \binom{d}{k} \exp(-\beta_k H),
\ee
where  $\beta_k = 2\beta k/d$  is discrete inverse temperature,  and $\mathcal Z_{\beta, d} = \Tr [e^{-\beta H}\cosh^d(\beta H/d)]$ is a modified partition function. In the limit of $d\to\infty$, the distribution over $\beta_k$ asymptotically approaches to the singular value at inverse temperature $\beta$.

To evaluate the performance at finite but large $d$, we evaluate the distance between the quasi-thermal distribution in Eq.\eqref{eqs:binom_gibbs} and the Gibbs state by taking into account only the leading terms in the small parameter $d^{-1}\ll 1$. In particular, we express
\be\label{eqs:simple_correction}
\begin{split}
\rho_{d} &\propto \exp(-\beta H)\cosh^d(\beta H/d) \\
&= \exp\Bigl(-\beta H + \frac{\beta^2 H^2}{2d}+O\left(d^{-2}\right)\Bigl).
\end{split}
\ee
Next, we establish the distance between this state and the Gibbs state by utilizing the following Lemma.

\begin{lemmaS} \label{lem:distance} 
Consider Gibbs states
\be \label{eqs:rho_modified}
\rho_\beta = \frac{1}{\mathcal Z_\beta}e^{-\beta H},\quad \rho'_\beta = \frac{1}{\mathcal Z'_\beta}e^{-\beta (H+\lambda V)},
\ee 
where $\mathcal Z_\beta = \Tr \{e^{-\beta H}\}$, $\mathcal Z'_\beta = \Tr \{e^{-\beta(H+\lambda V)}\}$, $H$ and $V$ are Hermitian operators, and $\lambda\ll 1$ is a small parameter. Then
\be\begin{split}
S\left(\rho_\beta\|\rho_\beta'\right) \leq \frac 12\lambda^2\beta^2\Bigl(\<V^2\>_\beta-\<V\>_\beta^2\Bigl)+O(\lambda^3),
\end{split}
\ee

where $S(\cdot||\cdot)$ is the relative entropy.
\end{lemmaS}
\textbf{Proof of Lemma \ref{lem:distance}.} We can express the distance between Eq.~\eqref{eqs:rho_modified} using the definition of relative entropy as
\be\begin{split}
S&\left(\rho_\beta\|\rho_\beta'\right) :=\Tr\bigl(\rho_\beta (\log\rho_\beta-\log\rho'_\beta)\bigl)=\<\log\rho_\beta-\log\rho_\beta'\>_\beta \\
&= -\log \mathcal Z_\beta+\lambda\beta\<V\>_\beta+\log[\Tr e^{-\beta (H+\lambda V)}].
\end{split}
\ee
By applying the Golden-Thompson inequality to the last term of this expression, we obtain
\be\label{eqs:s_GT_ineq}
\begin{split}
S\left(\rho_\beta\|\rho_\beta'\right)&\leq \lambda\beta\<V\>_\beta+\log \mathcal Z^{-1}\Tr e^{-\beta H}e^{-\lambda\beta V}\\
&= \lambda\beta\<V\>_\beta+\log\<e^{-\lambda\beta V}\>_\beta.
\end{split}
\ee
Lastly, we utilize the Taylor series expansion to expand the last term of the previous equation as
\be
\log\<e^{-\lambda\beta V}\>_\beta = -\lambda\beta\<V\>_\beta + \frac{1}{2}\lambda^2\beta^2\Bigl(\<V^2\>_\beta-\<V\>^2_\beta\Bigl)+O(\lambda^3).
\ee
By inserting this expansion into Eq.~\eqref{eqs:s_GT_ineq}, we arrive at the statement of the Lemma. \ensuremath{\square} \\

Using Eq.~\eqref{eqs:simple_correction} and Lemma 1, where we set $\lambda \sim d^{-1}$ and $V = \beta^2 H^2/2+O(d^{-1})$, we get
\be\label{eqs:S_average_m1}
\begin{split}
S(\rho_\beta\|\mathbb E[\rho_{\rm out}]) \leq  \frac{\beta^4}{8d^2}\Bigl(\<H^4\>_\beta-\<H^2\>^2_\beta\Bigl)+O\left(d^{-3}\right).
\end{split}
\ee
For physical local many-body Hamiltonians, the variance of the operator $H^2$ satisfies $\Var_\beta(H^2) = O(n^2)$, where $n$ is the number of qubits. Therefore, the algorithm converges if the number of iterations $d$ is of the order $d = O(\beta n)$.

\subsection{Divided Hamiltonian ($M>1$)}\label{s_sec:trotterized_rm_alg}

Let us prove a more general theorem that formalizes Theorem~\ref{thm:main_text_universal} in the main text. In our statement, we will specify the specific algorithm mode that produces the output for each quantity.

 \begin{theoremS}[formal of Theorem~\ref{thm:main_text_universal}] \label{thm:suppl_univ} Conisder Hamiltonian $H= \sum_{m=1}^M h_m$ and let $\xi : = \beta^2 M/d<1$ be small. Define $A:= \|h\|^2/4+O(\xi)$ for $h := M^{-1}\sum_{m=1}^M h_m^2$ and $C:= M^{-1}\sum_{m=1}^M{\rm Var}_\beta(h_m)\bigl(1+O(\xi)\bigl)$. Then for any $0<\epsilon\leq1$, the output of Algorithm 1 after $d$ cycles satisfies
\begin{align}\label{eq:theorem1_first_line}
\text{mode 1:}\quad&S(\rho_\beta\,\|\, \mathbb E[\rho_{\rm out}]) \leq  A\xi^2\;,\frac{}{}\\ 
\text{mode 2:}\quad&S(\rho_\beta\,\|\,\mathbb E_{\boldsymbol\theta}[\rho_{\rm out}(\boldsymbol\theta)]) \leq  C\xi\;,\\ 
\label{thrm1_second_line}
\text{mode 3:}\quad&{\rm Prob}\Bigl(S(\rho_\beta\,\|\,\rho_{\rm out}(\boldsymbol\theta)) \geq \frac{\xi C}{\epsilon }\Bigl)\leq \epsilon\;,
\end{align}
where $S(\rho\|\sigma)$ is the relative entropy and the expectation $\mathbb E[\rho_{\rm out}]$ is defined in Eq.~\eqref{eq:expect_avrg_rho}.\label{thm:1}
\end{theoremS}

We recall that the system state dynamics includes $d$ cycles, each consisting of $M$ gates in Eq.~\eqref{eq:rqc_gate}, followed by postselected measurements of the ancilla. Let $\rho_{km}(\boldsymbol \theta)$ be the state of the system qubits after the measurement of the $m$-th gate in the $k$-th cycle, where $\boldsymbol \theta:=\{\theta_{km}\}$ denotes the set of gate angles. We define $P_{km}(\boldsymbol\theta)$ to be the probability of acceptance of the overall circuit at this stage, i.e., that all ancillas up to this point are measured in the zero state. We then define $r_{km}(\boldsymbol \theta) := P_{km}(\boldsymbol \theta)\rho_{km}(\boldsymbol \theta)$, which for $m\geq 2$ and a given cycle $k$, can be written as
\be\begin{split}
r_{km}(\boldsymbol \theta) &= P_{k,m-1}(\boldsymbol \theta)\times\\
&\Tr_a\Bigl\{I_{\rm sys}\otimes|0\>\<0| U_{km}\Bigl(\rho_{k,m-1}(\boldsymbol \theta)\otimes |0\>\<0|\Bigl)U^\dag_{km}\Bigl\}\\
& = \Phi_m(r_{k,m-1}(\boldsymbol \theta))
\end{split}
\ee
where $U_{km}: = U(\theta_{km},h_{m})$ is defined in Eq.~\eqref{eq:rqc_gate} in the main text and we defined the linear map
\be
\Phi_m(\rho) := \cos(\theta_{km}\sqrt{\beta h_m/d})\rho\cos(\theta_{km}\sqrt{\beta h_m/d}).
\ee
 Using this recurrence relation, we can also connect the probability-weighted state of the system at the end of the $k-1$st cycle to the one at the beginning of the $k$th cycle
\be\label{eq:rec_condition}
r_{kM}(\boldsymbol \theta) = \Phi_{M}\circ\Phi_{M-1}\circ\dots \circ\Phi_{1}(r_{k-1,M}) 
\ee
 where $\mathcal A\circ \mathcal B(\rho) = \mathcal A(\mathcal B(\rho))$ is the composition of channels $A$ and $B$.
 
We set the input state to be the fully mixed state, i.e. $\rho_{0}=I/2^{n}$, where $I$ is $2^n\times 2^n$ identity matrix. Then, the overall circuit output is
\be\label{eqs:circuit_output}
P_{M,d}(\boldsymbol \theta)\rho_{M,d}(\boldsymbol \theta) = \mathcal C(\boldsymbol \theta)\rho_0\mathcal C^\dag(\boldsymbol \theta) = \frac 1{2^n}\mathcal C(\boldsymbol \theta)\mathcal C^\dag(\boldsymbol \theta),
\ee
where we defined the propagator
\be
\mathcal C(\boldsymbol \theta) := \prod_{k=1}^d\Biggl(\prod_{m=1}^M\cos\theta_{km}\sqrt{\beta h_{m}/d}\Biggl).
\ee
and $\prod_{m=1}^M$ is ordered direct product defined as $\prod_{m=1}^M a_m := a_M\dots a_1$.
The asymptotic behaviour of the output state in Eq.~\eqref{eqs:circuit_output} is given by the following Lemma.
\begin{lemmaS} \label{lem_asymptotic_C}
Assuming that $d^{-1}$ is small, then
\be
 \mathcal C(\boldsymbol \theta) \mathcal C^\dag(\boldsymbol \theta) = \exp\Bigl[-\beta \Bigl(H+\frac 1dV\Bigl)\Bigl],
 \ee
 where
 \be
 \begin{split}
 V = \sum_{m=1}^M\sum_{k=1}^d\Bigl[\bigl(\theta^2_{km}-1\bigl)h_m+&\frac{\beta}{6d}\theta^4_{km}h^2_m\Bigl]+O(1).
 \end{split}
\ee
\end{lemmaS}
\textbf{Proof of Lemma \ref{lem_asymptotic_C}}. First, we use Taylor's expansion by inverse depth $d^{-1}$ to express
\be
\begin{split}
\cos\theta_{km}&\sqrt{\beta  h_{m}/d} = 1-\frac{\beta h_{m}}{2d}\theta^2_{km}+\frac{\beta^2h_{m}^2}{24d^2}\theta_{km}^4+O\Bigl(\frac 1{d^3}\Bigl)\\
&\qquad = \exp\Biggl[-\frac{\beta h_{m}}{2d}\theta^2_{km}-\frac{\beta^2h_{mk}^2}{12d^2}\theta^4_{km}+O\Bigl(\frac 1{d^3}\Bigl)\Biggl] \\
&\qquad = \exp\Bigl(-\frac{\xi_{mk}}{2d}\Bigl),
\end{split}
\ee
where $\xi_{mk} := \beta\theta^2_{km}h_{m}+\beta^2\theta^4_{km}h_{mk}^2/6d+O(d^{-2})$. Using this representation, we rewrite
\be
\begin{split}
\mathcal C(\boldsymbol \theta) \mathcal C^\dag(\boldsymbol \theta) = \mathcal E(\boldsymbol \theta) \mathcal E^\dag(\boldsymbol \theta).
\end{split}
\ee
where
\be
\begin{split}
\mathcal E(\boldsymbol \theta) :=& \prod_{k=1}^d\prod_{m=1}^M \exp\Bigl(-\frac{\xi_{mk}}{2d}\Bigl).
\end{split}
\ee
Using second-order Suzuki-Trotter product formula (see \cite{childs2019,childs2021theory}), we express
\be
\begin{split}
\mathcal E(\boldsymbol \theta) \mathcal E^\dag(\boldsymbol \theta)=\exp\biggl(-\frac 1d\sum_{m=1}^M\sum_{k=1}^d\xi_{mk}+O\Bigl(\frac 1{d^2}\Bigl)\biggl).
\end{split}
\ee
Using this result, we obtain the expression

\be
\begin{split}
\mathcal C&(\boldsymbol \theta)\mathcal C^\dag(\boldsymbol \theta) = \\
&= \exp\Bigl(-\frac \beta d\sum_{m=1}^M\sum_{k=1}^d\Bigl\{\theta^2_{km}h_m+\frac{\beta}{6d}\theta^4_{km}h_{m}^2\Bigl\}+O\Bigl(\frac 1{d^2}\Bigl)\Bigl)\\
&= \exp\Bigl(-\beta H-\frac{\beta}{d}\sum_{m=1}^M\sum_{k=1}^d\Bigl\{(\theta^2_{km}-1\bigl)h_m\\
&\qquad\qquad\qquad\qquad\qquad +\frac{\beta}{6d}\theta^4_{km}h_{m}^2\Bigl\}+O\Bigl(\frac 1{d^2}\Bigl)\Bigl).
\end{split}
\ee
This leads us to the statement of the Lemma. \ensuremath{\square}\\

Next, let us derive the performance bounds of the algorithm in mode 1. For this purpose, we first express the expected density matrix as
\be
\mathbb E[\rho_{\rm out}] = \mathbb E_{\boldsymbol \theta} \mathcal C(\boldsymbol \theta) \mathcal C^\dag(\boldsymbol \theta)/\Tr \Bigl(\mathbb E_{\boldsymbol \theta}  \mathcal C(\boldsymbol \theta) \mathcal C^\dag(\boldsymbol \theta)\Bigl)
\ee
Using Lemma~\ref{lem_asymptotic_C}, we get
\be
\begin{split}
S(\rho_\beta\|\mathbb E[\rho_{\rm out}]) &= \<\rho_\beta-\mathbb E[\rho_{\rm out}]\>_\beta \\
&  =- \log \mathcal Z_\beta -\beta H+\log \mathbb E_{\boldsymbol \theta} \Tr e^{-\beta (H+V/d)}\\
&\quad -\<\log\mathbb E_{\boldsymbol \theta} e^{-\beta(H+V/d)}\>_\beta
\end{split}
\ee
where the operator $V$ is defined in Lemma~\ref{lem_asymptotic_C}. In order to evaluate this expression, we rewrite the exponent as
\be\label{eqs:9127a0s8f}
\exp(-\beta H-\beta V/d) = \exp(-\beta H)\exp(-\beta \varphi(V)/d),
\ee
where we introduced the linear transformation
\be\label{eqs:97tqw87dtvw86}
\varphi(O) := \int_0^1 d\tau e^{\tau H}O e^{-\tau H}.
\ee
Based on the decomposition in Eq.~\eqref{eqs:9127a0s8f}, we can rewrite relative entropy as
\be
S(\rho_\beta\|\rho_{\beta,d}) = \log \mathbb E_{\boldsymbol \theta} \<e^{-\beta \varphi(V)}\>_\beta -\<\log \mathbb E_{\boldsymbol \theta} e^{-\beta \varphi(V)}\>_\beta.
\ee
Next, we evaluate the leading order contribution using Taylor expansion as
\be
S(\rho_\beta\|\rho_{\beta,d}) = \frac{1}{2d^2}\Bigl(\<[\mathbb E_{\boldsymbol \theta} \varphi( V)]^2\>_\beta - \<\mathbb E_{\boldsymbol \theta} \varphi(V)\>_\beta^2\Bigl)+O(d^{-3}).
\ee
The expectation over the gate angles is
\be
\mathbb E_{\boldsymbol \theta} \varphi(V) = \frac{\beta^2}{2d}\varphi(h)+O(d^{-3}),
\ee
where we use the notation
\be
h := \frac 1M\sum_{m=1}^M h_m^2.
\ee
Using the property
 $\<\varphi (O)\>_\beta = \<O\>_\beta$, we finally express
\be
S(\rho_\beta\|\rho_{\beta,d}) = \frac{\beta^4M^2}{8d^2}\Bigl(\<\varphi(h)^2\>_\beta-\<h\>_\beta^2\Bigl) +O(d^{-3}).
\ee
This formula is a generalization of Eq.~\eqref{eqs:S_average_m1}.
Because the transformation in Eq.~\eqref{eqs:97tqw87dtvw86} does not change the spectral norm of the operator, we can bound the expression as
\be
S(\rho_\beta\|\rho_{\beta,d}) \leq A\frac{\beta^4M^2}{d^2}.
\ee
where $A = \|h\|^2/4+O(d^{-1})$. Assuming that the norm $\|h_m\|\leq O(1)$ is finite, parameter $A$ is also constant, $A \sim O(1)$.

Next, to derive the bounds for modes 2 and 3, we 
combine Lemmas \ref{lem:distance} and \ref{lem_asymptotic_C} to express
\be
\mathbb E_{\boldsymbol \theta} S(\rho_\beta\| \rho(\boldsymbol \theta))\leq \frac{\beta^2}{2d^2}\mathbb E_{\boldsymbol \theta}\Bigl(\<V^2\>_\beta-\<V\>^2_\beta\Bigl)+O(d^{-3}).
\ee
We obtain
\be
\mathbb E_{\boldsymbol \theta} \<V\>^2_\beta =2d\sum_{m=1}^M \<h_m\>^2_\beta+O(1).
\ee
Therefore, the expected relative entropy is
\be\label{eqs:re_bound}
\mathbb E_{\boldsymbol \theta} S(\rho_\beta\| \rho(\boldsymbol \theta))\leq \frac{C\beta^2M}{d},
\ee
where we denoted
\be
C = \frac{1}{M}\sum_{m=1}^M\Var_\beta(h_m)+O(d^{-1}).
\ee
where $\Var_\beta(x) = \<x^2\>_\beta-\<x\>^2_\beta$.

By Jensen's inequality, we have
\be
S(\rho_\beta\| \mathbb E[\rho_{\beta,d}]) \leq \mathbb E S(\rho_\beta\| \rho_{\beta,d}) \leq  \frac{C\beta^2M}{d}.
\ee
Using Markov's inequality, for any $x>0$ we have
\be
{\rm Prob}\Bigl(S(\rho_\beta\| \rho(\boldsymbol \theta))\geq x\Bigl)\leq \frac 1x\mathbb E_{\boldsymbol \theta} S(\rho_\beta\| \rho(\boldsymbol \theta)).
\ee
Using Eq.~\eqref{eqs:re_bound} and choosing $x = \mathbb E_{\boldsymbol \theta} S(\rho_\beta\| \rho(\boldsymbol \theta))/\epsilon$, we obtain the expression
\be
{\rm Prob}\Bigl(S(\rho_\beta\| \rho(\boldsymbol \theta))\geq \frac{C\beta^2 M}{\epsilon d}\Bigl)\leq \epsilon.
\ee
Thus, we prove Theorem 1. 

\begin{figure*}[t]
    \centering
  \includegraphics[width=1\linewidth]{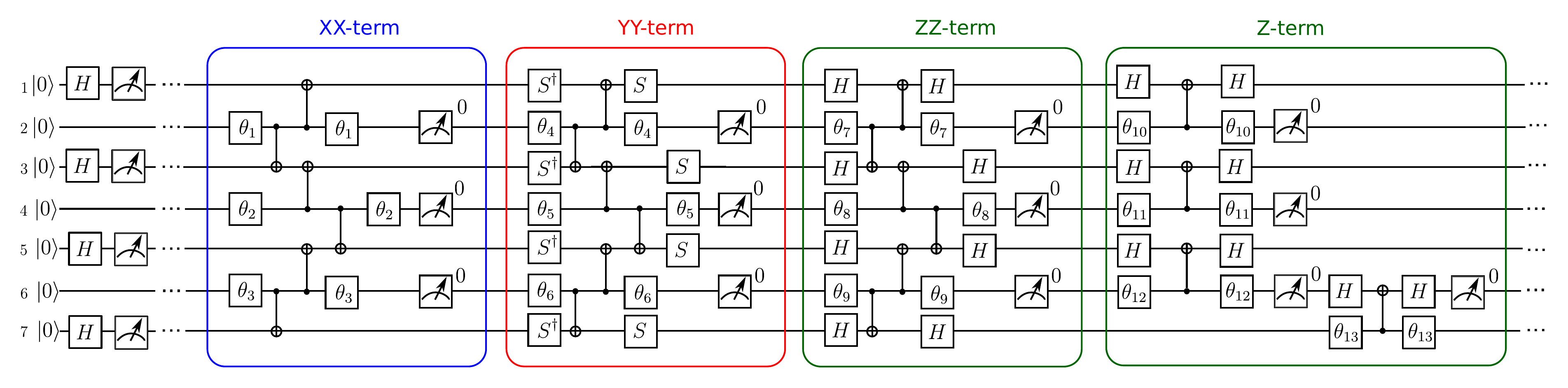}
  \caption{\textbf{Universal algorithm with native gates}. The circuit for the universal algorithm contains $d$ identical cycles; here we show only one cycle. In each cycle, three sub-cycles implement the terms $XX$, $YY$, $ZZ$, and $Z$, according to the decomposition in Eqs.~\eqref{eqs:two-qubit_gate_decomposition} and \eqref{eqs:three-qubit_gate_decomposition}. Gates labeled $H$ and $S$ denote Hadamard gates and standard $S$ gates, gates labeled $\theta_k$ (here we have omitted the cycle label) implement unitaries $u_{km} = \exp(\frac 12 i\theta_{k} v_m X) $, $v_m = \sqrt{\beta \alpha_m/d}$, $\alpha_m$ is the coefficient if front of the Hamiltonian Pauli term, and the symbol ``$0$'' next to a box indicates a postselected measurement.}
  \label{fig:fig1supp}
\end{figure*}

\subsection{Success probability}\label{Supp1:C}

Our final goal is to derive the acceptance probability, i.e. the probability that all the ancilla measurements return zero. The probability of such event success is given by \eqref{eqs:circuit_output} as
\be
\begin{split}
p \equiv P_{M,d}(\boldsymbol \theta) &= \frac 1{2^n}\mathbb E_{\boldsymbol \theta} \Tr \mathcal C(\boldsymbol \theta)\mathcal C^\dag(\boldsymbol \theta)\\
&=2^{-n}\Tr e^{-\beta H}+O(d^{-1})
\end{split}
\ee
Therefore, for large-$d$ circuits 
\be\label{eqs:succ_prob}
P \simeq 2^{-\alpha n}, \quad \alpha = 1-(\log_2 \mathcal Z_\beta)/n
\ee
Note that $\beta = 0$ corresponds to the partition function  $\mathcal Z_\beta = 2^n$ for any Hamiltonian. In this case the success probability is $P_d=1$. This observation is easily explained by noting that in the case $\beta = 0$ all the gates in the circuit are trivial and the system state is already initialized in infinite-temperature (fully mixed) state.

\subsection{Resource estimation}\label{Supp2:ResourceEstim_Univ}

In this section, we provide a summary of the resources required for implementing the algorithm. As in the main text, $n$ is the number of qubits for the target Hamiltonian, $\beta$ is the inverse temperature, and $\epsilon$ is the algorithm's error.
\begin{table}[h!]
\begin{tabular}{|l|c|}
\hline
Type of Hamiltonians & any                                                                                                                      \\ \hline
Performance time     & $2^{\Theta(n)}$                                                                                                          \\ \hline
Min \# of ancilla    & $1$                                                                                                                   \\ \hline
Min depth      & \begin{tabular}[c]{@{}c@{}}$O(\beta^2n^2/\epsilon)$ (generic) \\  $O(\beta^2n^2/\sqrt{\epsilon})$(average) \end{tabular} \\ \hline
\end{tabular}
\end{table}

We provide the number of gates both for the circuit-averaged sampling and for a generic circuit outcome.

\subsection{Details of simulation}\label{Supp1:E}

We simulate the \textit{accepted} output state using the expression
\be\label{eqs:rho_out_universal}
\rho_{\rm out}(\boldsymbol \theta) = \frac{1}{P(\boldsymbol \theta)}\mathcal E'_{d} \circ \dots \circ \mathcal E'_{1}(\rho_{0}),
\ee
where $\rho_0$ is the initial fully mixed state of the system qubits, $\boldsymbol\theta = \{\theta_{km}\}$ are gate angles, and $P(\boldsymbol \theta)$ is the probability of accepting the circuit,
\be\label{eqs:P_universal}
P(\boldsymbol \theta) = \Tr \mathcal E'_{d} \circ \dots \circ \mathcal E'_{1}(\rho_{0}).
\ee
The channels $\mathcal E'_{k}$ represent the transformation of the system qubits during the corresponding algorithm cycle. Such channels are the composition of channels generated by each gate in Eq.~\eqref{eq:rqc_gate} and corresponding ancilla measurement, i.e.
\be
\mathcal E'_{k} = \mathcal E'_{kM}\circ\dots\circ\mathcal E'_{k1},
\ee
where
\be
\mathcal E'_{km}(\rho) := \Tr_a\Bigl\{(I_{\rm sys}\otimes|0\>\<0|)U_{km}\bigl(\rho\otimes |0\>\<0|\bigl)U^\dag_{km}\Bigl\},
\ee
$U_{km} = U(\theta_{km},h_m)$, and $|0\>$ represents the initial state of the ancilla qubit.
The probability of sampling the eigenstate of the Hamiltonian is given by the formula
\be
n_\mu =  \mathbb E_{\boldsymbol \theta} P(\boldsymbol \theta)\<\mu|\rho_{\rm out}(\boldsymbol \theta)|\mu\>/\mathbb E_{\boldsymbol \theta} P(\boldsymbol \theta),
\ee
where $|\mu\>$ are eigenstates of the Hamiltonian.
 The result is shown in Fig.~\ref{fig:fig2}(a) right panel (blue circled lines) for $\beta = 1$ compared to the theoretical prediction (dashed line)
\be
n^{\rm theor}_\mu = \mathcal Z^{-1}\exp(-\beta E_\mu), \quad \mathcal Z = \sum_\mu \exp(-\beta E_\mu).
\ee
We also evaluate the temperature dependence of the output energy
\be
E =  \mathbb E_{\boldsymbol \theta} P(\boldsymbol \theta)\Tr(\rho_{\rm out}(\boldsymbol \theta) H)/\mathbb E_{\boldsymbol \theta} P(\boldsymbol \theta),
\ee
and compare it to theoretical thermal value,
\be
E_{\rm theor} = \sum_\mu  E_\mu \exp(-\beta E_\mu).
\ee
The output energy dependence is shown in Fig.~\ref{fig:fig2}(b) right panel by blue line (circles) and the thermal value is depicted by dashed line.

To consider the effect of noise, we replace the channels in Eqs.~\eqref{eqs:rho_out_universal} and Eq.~\eqref{eqs:P_universal} by
\be
\mathcal E'_{km}\to \mathcal E'^{\rm noisy}_{km}:= \mathcal D_{mi_q}\circ\dots\circ\mathcal D_{mi_1}\circ\mathcal E'_{km}
\ee
where $i_1\dots i_q$ enumerate the qubits on which $U_{km}$ is acting, and we defined single-qubit depolarizing channel as
\be\label{eqs:noisy_transform2}
\begin{split}
\mathcal D_{mi}(\rho) =&\bigl(1-3p_m\bigl)\rho + p_m\Bigl(X_i\rho X_i+Y_i\rho Y_i+Z_i\rho Z_i\Bigl),
\end{split}
\ee
$X_i$, $Y_i$, and $Z_i$ are Pauli operators acting on qubit $i$, and $p_m$ is the depolarizing error probability for gate $m$. The error probability is determined by the gate's structure. We demonstrate below that any two-qubit $U_{km}$ for Hamiltonian Pauli terms can be decomposed using single-qubit gates and one two-qubit native gate (for example, CNOT). In addition, any such three-qubit gate $U_{km}$ can be decomposed using two CNOTs. Given that the CNOTs contribute the majority of error, it is reasonable to assume that three-qubit gates will have twice the error of two-qubit gates for Pauli Hamiltonian terms. Fig.~\ref{fig:fig2}. Right panels of Figures (a) and (b) show the results of the noisy simulation as a green curve (diamonds) .

\subsection{Optimized implementation}\label{Supp1:F}

The representation of generic gates as standard one- and two-qubit gates is resource-demanding. There is a way to represent a generic two-qubit gate using only three two-qubit entangling gates \cite{zhang2003exact}. A generic three-qubit gate, however, would require many two-qubit entangling gates to be implemented. 

In our case, there is a way to avoid similar computational burdens. As we recall, ancilla qubits are measured after every gate, and we accept circuits if the measurement is zero. Therefore, each two- or three-qubit gate has a zero ancilla value in its input. Then, in the case of this particular input and Pauli terms in Eq.~\eqref{eq:hamiltonian}, there is a simpler method of implementing desired circuits. Gates dedicated to the terms $h_m\propto Z_i$ can be implemented, for example, by just a single entangling gate
\be\label{eqs:two-qubit_gate_decomposition}
\begin{split}
& {\rm R}_X(\theta)|\psi\>|0\> \equiv\\
&\qquad{\rm R}_X(\theta)H_i \cdot {\rm CNOT}_{A\to i}\cdot H_i{\rm R}_X(\theta)\,|\psi\>|0\>,
\end{split}
\ee
where $H_i$ is Hadamard gate for qubit $i$, ${\rm CNOT}_{A\to i}$ is a controlled NOT gate that applies to qubit $i$ and uses the ancilla $A$ as control, and ${\rm R}_X(\theta) = \exp(\frac 12 i\theta X_A)$ acts on ancilla qubit. Similarly, the gate implementing the term $h_m \propto X_iX_j$ requires only two controlled-NOT operations and single-qubit gates,
\be\label{eqs:three-qubit_gate_decomposition}
\begin{split}
&\exp\Bigl(i\theta \frac 12(I+X_iX_j) X_A\Bigl)|\psi\>|0\> \equiv\\
& \quad{\rm R}_X(\theta) \, {\rm CNOT}_{A\to i} \, {\rm CNOT}_{A\to j}\,{\rm R}_X(\theta)|\psi\>|0\>,\\
&\exp\Bigl(i\theta \frac 12(I+Y_iY_j) X_A\Bigl)|\psi\>|0\> \equiv\\
& \quad{\rm R}_X(\theta) \, S_i\, S_j\, {\rm CNOT}_{A\to i} \, {\rm CNOT}_{A\to j}\, S^\dag_i\, S^\dag_j\,{\rm R}_X(\theta)|\psi\>|0\>,\\
&\exp\Bigl(i\theta \frac 12(I+Z_iZ_j) X_A\Bigl)|\psi\>|0\> \equiv\\
& \quad{\rm R}_X(\theta) \,H_i\, H_j\, {\rm CNOT}_{A\to i} \, {\rm CNOT}_{A\to j}\, H_i\, H_j\, {\rm R}_X(\theta)|\psi\>|0\>,
\end{split}
\ee
where $S_i$ is $S$-gate on $i$th qubit.
The circuit for the Hamiltonian in Eq.~\eqref{eq:hamiltonian} is shown in Fig.~\ref{fig:fig1supp}. As this circuit is built from native gates, its implementation is straightforward using tools such as IBM Qiskit.

\begin{figure}[t]
    \centering
  \includegraphics[width=1\linewidth]{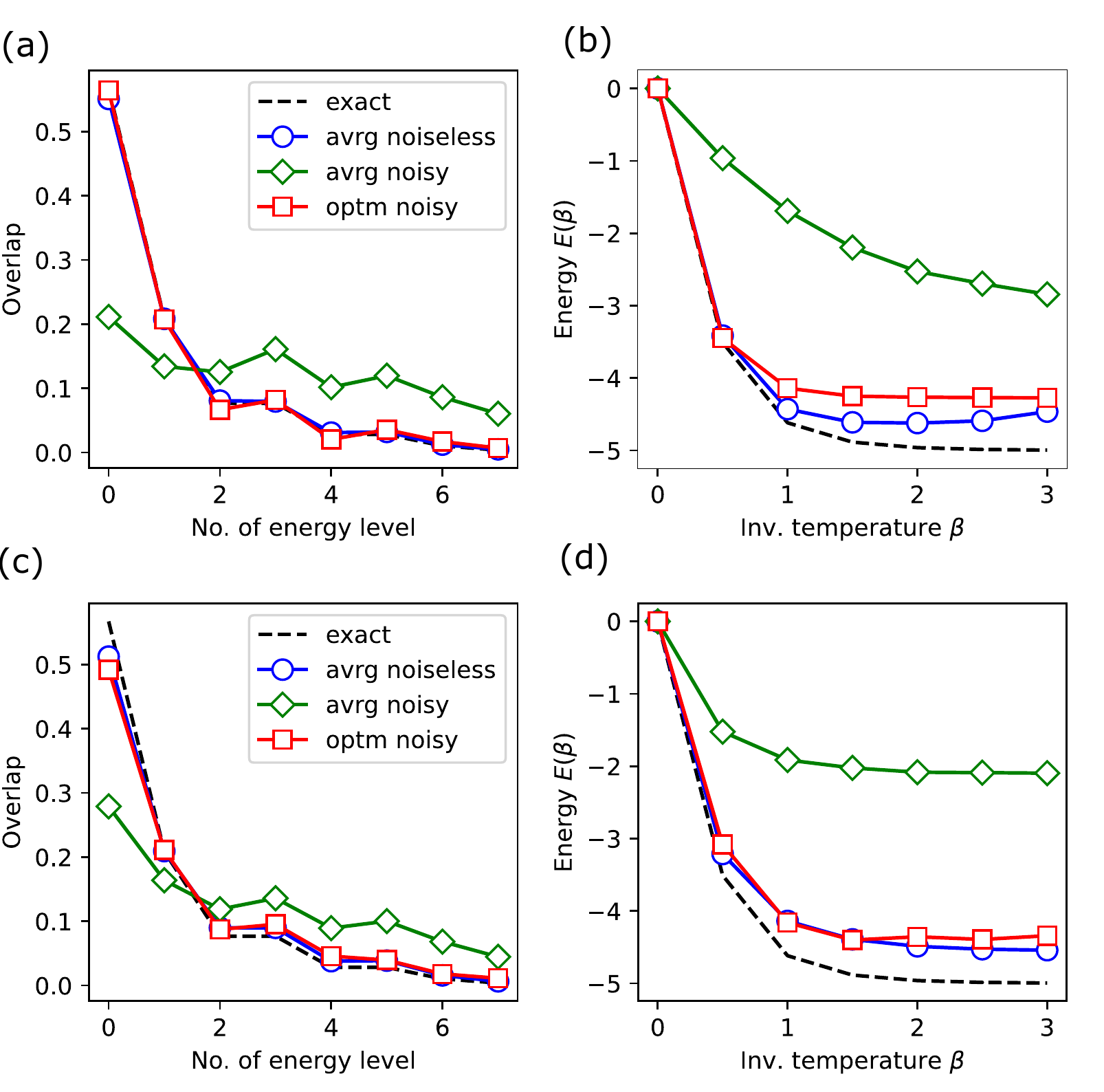}
  \caption{
\textbf{Additional simulations.} (a)-(b) Simulation results for universal algorithm utilizing $n_a = 2$ ancilla qubits for 3-qubit Hamiltonian in Eq.~\eqref{eqs:1d_qubit_system}, $t = -2$, $U=4$ and $h_i = -1$ (Heisenberg model). The inverse temperature is $\beta = 1$ and number of cycles is $d=5$. Panel (a) shows the overlap between expected output state in Eq.~\eqref{eq:expect_avrg_rho} and the eigenstates of the Hamiltonian for noiseless circuit (blue circles), noisy circuit (green diamonds), noisy optimized circuit (red squares), and exact solution (dashed line). We use a depolarizing noise model with single-qubit error probability $p_2=0.01$ for 2-qubit gates and $p_3=0.02$ for 3-qubit gates. An averaged circuit output is computed for $10^3$ successful sample runs. (b) Output state energy as a function of temperature for the same setting as the panel (a). (c)-(d) Performance for ergodic algorithm for the same setting as Panel (a)-(b), except using $d=20$ cycles, $\gamma = 0.1$, and single-qubit depolarizing error probability $p = \Gamma t$ after each cycle, where $t$ is the cycle time and $\Gamma = 10^{-3}$. The average is based on $10^3$ sample runs.}
  \label{fig:fig3s}
\end{figure}
\section{Mapping to qubits}\label{s_sec:fh_mapping}

In the main text, we considered the hard-core boson model Hamiltonian, which is expressed as
\be
H = -J\sum_{\<i,j\>} \Bigl(a^\dag_ia_{j}+a^\dag_ja_{j}\Bigl)+U\sum_{\<i,j\>}n_in_{j},
\ee
where $a_i$, $a_i^\dag$ are hardcore-boso Fock operators that satisfy conditions $[a_j,a^\dag_j] = 0$ for all $i\neq j$ and $\{a_i,a^\dag_i\}=1$ for the same site. These operators can be mapped to the qubit Pauli operators using the transformation
\be
\begin{split}
&a_i = \frac 12(X_i-iY_i), \qquad a^\dag_i = \frac 12(X_i+iY_i).
\end{split}
\ee
For the one-dimensional setting, we obtain the qubit Hamiltonian
\be\label{eqs:1d_qubit_system}
\begin{split}
H = -\frac J2\sum_{i=1}^{N-1}\Bigl(&X_iX_{i+1}+Y_{i}Y_{i+1}\Bigl)
\\&+\frac U4\sum_{i=1}^{N-1}Z_iZ_{i+1}
+\sum_i h_iZ_i,
\end{split}
\ee
where $h_i = -U/2$ for $j=2,\dots,N-1$ and $h_i = -U/4$ for $j=1,N$. It is possible to generate the Gibbs state for this Hamiltonian using this one-dimensional circuit architecture. Some additional simulation data for the spin model in Eq.~\eqref{eqs:1d_qubit_system} are shown in Fig.~\ref{fig:fig3s}.

\end{document}